\documentclass[a4paper,11pt]{article}
\usepackage{jheppub} % for details on the use of the package, please 
\usepackage[T1]{fontenc} % if needed
\usepackage{enumitem}
\usepackage{xcolor} 
\usepackage[utf8]{inputenc}
\usepackage{cleveref}
\usepackage{amsmath}
\usepackage{amssymb}
\usepackage{pifont}
\usepackage{bm}
\usepackage{enumitem}
\usepackage{makecell}
\usepackage{subfig}
\usepackage{tabularx,booktabs}
\newcolumntype{C}{>{\centering\arraybackslash}X}
\usepackage{graphicx}
\usepackage{multirow}
\usepackage[normalem]{ulem}
\usepackage{comment}
\usepackage{booktabs}
\usepackage{slashed}
\usepackage{hyphenat} 
\usepackage{multirow,bigdelim}  
\usepackage{lipsum}
\usepackage{makecell}

\usepackage{tikz-feynman}
\tikzfeynmanset{compat=1.1.0}
\setlist[enumerate]{wide=0pt, widest=99,leftmargin=\parindent, labelsep=*}

\newdimen\arrowsize
\newdimen\mylw
\pgfkeys{/my arrows/chemeq/.style={draw,thick,double distance=2pt,onearc-onearc}}
\pgfkeys{/my arrows/size/.code={\pgfsetarrowoptions{onearc}{#1}}}
\def\myalw{.4pt}
\pgfarrowsdeclare{onearc}{onearc}{%
  \mylw=\myalw
  \pgfarrowsleftextend{-\pgfgetarrowoptions{onearc}-.5\mylw}
  \pgfarrowsrightextend{0pt}
}{%
  \pgfsetdash{}{0pt}
  \mylw=\pgflinewidth
  \pgfsetlinewidth{\myalw}
  \advance\arrowsize by.5\pgflinewidth
  \pgfpathmoveto{\pgfpoint{-\pgfgetarrowoptions{onearc}}{-\pgfgetarrowoptions{onearc}-.5\mylw}}%
  \pgfpatharc{180}{90}{\pgfgetarrowoptions{onearc}}
  \pgfusepathqstroke
}

\title{\boldmath  {
	  A Roadmap for neutrino charge assignments in $U(2)_F$ Flavor Models: Implications for LFV processes and leptonic anomalous magnetic moments}}
\author[a,b]{Alessio Giarnetti,}
\author[a,b,c]{Simone Marciano,}
\author[a,b]{Davide Meloni,}
\author[a,b]{and Mirko Rettaroli}

\affiliation[a]{Dipartimento di Matematica e Fisica, Universit\'a di Roma Tre,
Via della Vasca Navale 84, 00146, Roma, Italy}

\affiliation[b]{INFN Sezione di Roma Tre, Via della Vasca Navale 84, 00146, Roma, Italy}

\affiliation[c]{Instituto de Física Corpuscular (IFIC), Universitat de València-CSIC, 
Parc Científic UV, C/Catedrático José Beltrán, 2, E-46980 Paterna, Spain}

\emailAdd{alessio.giarnetti@uniroma3.it}
\emailAdd{simone.marciano@ific.uv.es}
\emailAdd{davide.meloni@uniroma3.it}
\emailAdd{mirko.rettaroli@uniroma3.it}

\abstract{
We build upon a simple $U(2)_F$ model of flavor, in which all fermion masses and mixing hierarchies arise from powers of two small parameters controlling $U(2)_F$ breaking. In the original formulation, an isomorphism to the discrete $D_6\times U(1)_F$ symmetry was invoked to generate a Majorana neutrino mass term. Here, we retain the successful features of that model for the charged leptons and quarks, while exploring alternative neutrino charge assignments within the $U(2)_F$ framework. This approach allows us to generate Majorana neutrino masses via the see-saw mechanism without introducing any additional symmetries nor invoking any isomorphism. We further examine the implications of our models for Lepton Flavor Violating (LFV) decays, analyzing the processes $\mu\rightarrow e\gamma$, $\tau\rightarrow\mu\gamma$ and $\tau\rightarrow e\gamma$ 
and their connection with the leptonic anomalous magnetic moments. 
We show that within the Standard Model Effective Field Theory (SMEFT) approach the current limits on the branching ratios of $\mu\rightarrow e \gamma$ LFV decays obtained in our $U(2)_F$ models are not compatible with the central value of the recent measurement of the $(g-2)_\mu$, thereby suggesting that either $(g-2)_\mu$ must be very close to the Standard Model predictions, as the latest experimental and theoretical results seem to suggest, or the invoked flavor symmetry is not appropriate to describe an anomalous muon magnetic moment.} 


\makeatletter
\gdef\@fpheader{Prepared for submission to JHEP}
\makeatother

\begin{document} 
\maketitle
\flushbottom

\section{Introduction} \label{sec:intro}

In the last two decades, neutrino physics has undergone a strong revolution. The reason is related to the discovery of neutrino oscillations \cite{Super-Kamiokande:1998kpq}, which are described by 6 real parameters: 2 mass squared differences $\Delta m_{\text{sol}}^2$ and $\Delta m_{\text{atm}}^2$, 3 mixing angles $\theta_{12}$, $\theta_{23}$ and $\theta_{13}$, and a CP violating phase $\delta_\text{CP}$. 
The presence of two mass squared differences implies that at least two of the three neutrino masses are nonzero, in contrast with the original formulation of the Standard Model (SM) where neutrinos were assumed massless \cite{Gaillard:1998ui}.

Despite the increasingly precise knowledge of the oscillation parameters, several aspects of neutrino physics remain unknown. From the experimental point of view, current oscillation experiments could not be able to measure the absolute neutrino mass scale and present a limited capability of discerning a possible CP violation in the lepton sector as well as the neutrino mass ordering. From the theoretical point of view, there are no fully satisfactory theories which explain why neutrinos are massive and why their masses are so smaller compared to those of the other fermios. Moreover, the true nature of neutrinos, namely whether they are Dirac or Majorana particles~\cite{Majorana:1937vz} remains unknown.
The neutrino mixing pattern makes also difficult to find a connection between quark and lepton properties; this is a part of the well-known \textit{flavor problem}~\cite{King:2003xn,Feruglio:2015jfa,Abbas:2023ivi}. Indeed a different flavor mixing structure between quarks and leptons clearly emerges if we compare the entries of $V_{\text{CKM}}$ (the Cabibbo-Kobayashi-Maskawa matrix~\cite{Cabibbo:1963yz,Kobayashi:1973fv}) and $U_{\text{PMNS}}$ (the Pontecorvo-Maki-Nakagawa-Sakata matrix~\cite{Pontecorvo:1957cp,Pontecorvo:1957qd,Maki:1962mu,Pontecorvo:1967fh}): the former is almost diagonal, while the entries of the latter are almost all comparable to each other.

The search for an answer to all these questions and the need to give a flavor structure to the SM particles led to the proposal of various flavor symmetries, which constitute an extension to the SM gauge group $SU(3)_C \otimes SU(2)_L \otimes U(1)_Y$ and under which the three particle families transform in the same way.
The general idea is the following: introduce one or more new scalar fields, which are neutral under the SM but charged under the
chosen flavor symmetry. These fields, known as \emph{flavons}, transform according to some irreducible
representation of the flavor symmetry group. Flavons interact with leptons and the Higgs field
according to the most general flavor-invariant Lagrangian, which is built considering all matter
fields assigned to specific irreducible representations of the flavor group. The crucial point is that
the flavons interact diﬀerently with various leptons, meaning the interactions are not universal. This
non-uniform coupling allows for the construction of an invariant Lagrangian that, after flavor and gauge symmetry breaking, gives rise to the observed mixing matrix and mass spectrum.

Early oscillation data were well-compatible with a highly symmetric lepton mixing pattern, the \textit{Tri-Bi-Maximal} (TB) one~\cite{Harrison:2002er,Harrison:2002kp,Xing:2002sw,Harrison:2002et,Harrison:2003aw}, which implies $\sin^2\theta_{12} = 1/3$, $\sin^2\theta_{23} = 1/2$ and $\theta_{13} = 0$ and could be realized by non-abelian discrete symmetry (NADS) groups. Among the simplest groups we find $S_3$~\cite{Feruglio:2007hi}, $A_4$~\cite{%Ma:2001dn,Babu:2002dz,
Altarelli:2005yp}, and $S_4$~\cite{Bazzocchi:2009pv%,Grimus:2009pg
}.
However, since the discovery that the reactor mixing angle $\theta_{13}$ is nonzero \cite{DayaBay:2012fng}, this approach has gradually lost interest, as these models require (large) deviations from the leading-order predictions in order to be successful \cite{Giarnetti:2024vgs}— deviations that, we emphasize, can be nonetheless obtained from next-to-leading order corrections or by taking into account corrections from the charged lepton sector.

Recently~\cite{Feruglio:2017spp}, the NADS has regained popularity due to intriguing work on modular symmetries. These have been extensively studied within the context of string theory~\cite{Feruglio:2019ybq}. Interestingly, Feruglio demonstrated that it is always possible to find an isomorphism from the modular finite group $\Gamma_N$ of level $N\in \mathbb{N}$ to a non-abelian discrete symmetry group. In particular, for $N\leq 5$ the modular finite groups are isomorphic to $S_3,A_4,S_4$ and $A_5$. These models have attracted considerable attention, among other reasons, because they involve few free parameters, they do not generally require the presence of scalar fields like the flavons and are, therefore, more minimal and predictive. A list of few examples of such models follows: $\Gamma_2\cong S_3$~\cite{Kobayashi:2018vbk,Marciano:2024nwm}, $\Gamma_3\cong A_4$~\cite{Kobayashi:2018wkl,Criado:2018thu,Kobayashi:2018scp,Okada:2018yrn,Okada:2019uoy,Ding:2019zxk,Asaka:2019vev,Okada:2021qdf,Nomura:2022mgf,Devi:2023vpe}, $\Gamma_4\cong S_4$~\cite{Ding:2019gof,Penedo:2018nmg,Novichkov:2018ovf,deMedeirosVarzielas:2019cyj} and $\Gamma_5\cong A_5$~\cite{Criado:2019tzk,Novichkov:2018nkm,Ding:2019xna}.
Besides the NADS, alternative possibilities for addressing the neutrino mass and mixing problem based on continuous symmetries can be explored. The most minimal formulation relies on the addition of an abelian $U(1)_F$ symmetry group~\cite{Leurer:1992wg,Leurer:1993gy,Dreiner:2003hw,Petcov:1982ya,Arcadi:2022ojj}, whose main critical issue is its limited predictivity due to the large number of free parameters. For this reason, one can consider more predictive, though less minimal, models — such as those based on a $U(2)_F$ flavor symmetry~\cite{Linster:2018avp,Falkowski:2015zwa,Barbieri:1995uv,Barbieri:1997tu,Dudas:2013pja}.
The original interest on these models was certainly related to their ability to explain the values of quark masses and mixing, as well as to provide the absence of large Flavor Changing Neutral Currents (FCNCs) in the presence of low-energy Supersymmetry~\cite{Barbieri:1995uv,Barbieri:1997tu}. But it has been also shown that, once extended to the neutrino sector, the $U(2)_F$ symmetry is able to adjust the values of the lepton mixing parameters. In addition, the  New Physics (NP) interactions can induce Lepton Flavor Violating (LFV) decays and new contributions to the muon magnetic moment $(g-2)_\mu$~\cite{Muong-2:2023cdq,Muong-2:2021ojo,Muong-2:2006rrc,Aliberti:2025beg,Muong-2:2025xyk}. This raises the interesting question of whether the same NP that explains neutrino masses and mixing could also mediate LFV processes — and under what constraints — and whether it might also be responsible for the anomalous magnetic moment of the muon.
Numerous studies in the literature have explored this direction. For example, the electron and muon magnetic moments have been thoroughly analyzed within the framework of the Standard Model Effective Field Theory (SMEFT)~\cite{Buchmuller:1985jz,Grzadkowski:2010es,Alonso:2013hga}, where the new physics is assumed to lie above the Electroweak (EW) scale~\cite{Panico:2018hal,Aebischer:2021uvt,Allwicher:2021rtd,Kley:2021yhn}. Other works investigate the connection between the muon's anomalous magnetic moment $(g-2)_\mu$ and LFV processes, including scenarios involving modular flavor symmetries~\cite{Kobayashi:2021pav,Kobayashi:2022jvy}, $U(2)_{L_L}\otimes U(2)_{E_R}$ flavor model~\cite{Tanimoto:2023hse}, extended scalar sectors~\cite{Chowdhury:2015sla,DeJesus:2020yqx,Esch:2016jyx,Giarnetti:2023dcr,Giarnetti:2024psj} and radiative lepton mass models~\cite{Nakai:2025dmp}.

Driven by these motivations,
we study a non-supersymmetric $U(2)_F$ model of flavor, which is inspired by the analyses performed in~\cite{Linster:2018avp} where the proposed model has been shown to reproduce the quarks and leptons observables. 
The authors, after attempting to apply the $U(2)_F$ see-saw model to neutrinos, introduced a $D_6\times U(1)_F$ flavor symmetry with the goal of getting Majorana neutrino masses from the Weinberg operator with no fine-tuning. Instead, we provide see-saw Majorana neutrino mass terms solely within $U(2)_F$ by relaxing their charge requirements and making use of the arbitrariness in the assignment of the $U(2)_F$ quantum numbers to the right-handed (RH) neutrinos. We also show that the model discussed in~\cite{Linster:2018avp}, even by employing a $U(2)_F$ see-saw mechanism, can still successfully reproduce neutrino observables by lowering the combination of a few free parameters.

The paper is organized as follows: in Sec.~\ref{sec:model}, we describe the relevant features of our constructions.
In Sec.~\ref{sec:neutrinosector} we provide a catalog of possible neutrino mass matrices, obtained via the type-I see-saw mechanism.
Subsequently we derive predictions on the physical observables $m_{\beta}$ and $m_{\beta\beta}$.
In Sec.~\ref{LFVintr}, we analyse the LFV decays in light of the newest evaluation of the $(g-2)_\mu$ anomalous magnetic moment~\cite{Muong-2:2025xyk,Aliberti:2025beg} and we discuss our results in the Conclusions, Sec.~\ref{conclusions}.

\section{A realistic $U(2)_F$ model of flavor} \label{sec:model}

The first $U(2)_F$ models have been proposed in~\cite{Barbieri:1995uv,Barbieri:1997tu} in the context of SUSY, with flavor quantum numbers compatible with a unified SO(10) gauge group. In that case, the special structure of quark mass matrices lead to the prediction $V_{ub}/V_{cb} = m_u/m_c$ (with $V \equiv V_{CKM}$), which is strongly disfavored by the data.
However, it was shown that a viable $U(2)_F$ model with flavor quantum numbers compatible only with an unified $SU(5)$ gauge group was possible~\cite{Dudas:2013pja,Falkowski:2015zwa}. This class of models does not require the presence of SUSY.

In this paper we will follow the latter approach;
the breaking of the $U(2)_F$ flavor symmetry is described by two flavons $\phi$ and $\chi$, whose VEVs determine the hierarchical structure of the mass matrices in the various fermionic sectors of the SM.

\subsection{The charged lepton sector} \label{sec:fitquark}

We work with a global $U(2)_F$ symmetry, which is broken slightly below a cutoff scale $\Lambda$, corresponding to the mass scale of the new degrees of freedom. This mass scale must be very high to ensure that such additional dynamics do not affect the low-energy phenomenology we are interested in.
It can be demonstrated that $U(2)_F$ is locally isomorphic to $SU(2)_F \times U(1)_F$. 
Here we briefly revise the sucessfull construction described in \cite{Linster:2018avp}, 
whose field quantum numbers are the good ones to reproduce the quarks masses and CKM mixing as well as the ratio of the charged lepton masses, see Table~\ref{tab:quantnum}, \cite{Linster:2018avp,Falkowski:2015zwa}.
\renewcommand{\arraystretch}{1.5}
\begin{table}[h]
\centering
\begin{tabularx}{0.7\textwidth}{C|cc}
\Xhline{1.3pt} \rule[-0.25cm]{0pt}{0.7cm}
\bf{Fields / Representations} & $\mathbf{SU(2)_F}$ & $\mathbf{U(1)_F}$ \\ 
\Xhline{1pt}
$L_a$ , \ $D_a$ , \ $Q_a$ , \ $U_a$ , \ $E_a$ \hspace{0.4cm} ($a=1,2$)
 & $2$ & \phantom{+}$1$ \\ %\hline

%\hline \rule[-0.3cm]{0pt}{0.8cm}

$Q_3$ , \ $U_3$ , \ $E_3$ & $1$ & \phantom{+}$0$ \\ %\hline

%\hline \rule[-0.3cm]{0pt}{0.8cm}

$L_3$ , \ $D_3$ & $1$ & \phantom{+}$1$ \\ %\Xhline{1.3pt}
\hline
%\hline \rule[-0.7cm]{0pt}{1.6cm}
$\phi_a$ \hspace{0.3cm} ($a=1,2$) & $2$ & $-1$ \\ %\hline
%\hline \rule[-0.3cm]{0pt}{0.8cm}
$\chi$ & $1$ & $-1$ \\

\Xhline{1.3pt}

\end{tabularx}
\caption{\label{tab:quantnum} $SU(2)_F$ and $U(1)_F$ quantum numbers of SM fermions and flavons. The SM fermions are organized into representations of the Standard Model gauge group. The Higgs is a singlet under both $SU(2)_F$ and $U(1)_F$.}
\end{table}
\renewcommand{\arraystretch}{1}\noindent

Here, $L$ and $Q$ contain  the SM left-handed lepton doublet and quark doublet, respectively, while $E$, $D$ and $U$ are  the SM right-handed electron, down quark and up quark.
Each representation has three generations corresponding to the three flavors. As we can see from Table~\ref{tab:quantnum}, the first two generations transform as $SU(2)_F$ doublets and the third generations as singlets.
It is worth specifying that $SU(2)_F$ doublets and singlets should not be confused with the $SU(2)_L$ isospin doublets and singlets of the SM.
Regarding the scalar fields, the Higgs ($H$) is a singlet under both $SU(2)_F$ and $U(1)_F$, while the flavons $\phi$ and $\chi$ are an $SU(2)_F$ doublet and singlet, respectively.
This charge assignments are compatible with an $SU(5)$ GUT structure~\cite{Georgi:1974sy}, on which we will not further elaborate in this paper.

Since fermions are charged under $U(2)_F$, constructing invariant terms in the Lagrangian necessitates additional flavon insertions. This leads to the following form of the charged lepton Lagrangian:
\begin{equation} \label{eq:lage}
    \begin{split}
    \mathcal{L}_e &= \frac{e_{11}}{\Lambda^6} \chi^4 (\widetilde{\phi} L)_1 (\widetilde{\phi} E)_1 H + \frac{e_{12}}{\Lambda^2} \chi^2 (L E)_1 H + \frac{e_{13}}{\Lambda^3} \chi^2 (\widetilde{\phi} L)_1 E_3 H \ + \\
    &+ \frac{e_{22}}{\Lambda^2} (\phi L)_1 (\phi E)_1 H + \frac{e_{23}}{\Lambda} (\phi L)_1 E_3 H + \frac{e_{31}}{\Lambda^4} \chi^3 L_3 (\widetilde{\phi} E)_1 H \ + \\
    &+ \frac{e_{32}}{\Lambda^2} \chi L_3 (\phi E)_1 H + \frac{e_{33}}{\Lambda} \chi L_3 E_3 H \ ,
\end{split}
\end{equation}
where $\widetilde{\phi} = i \sigma_2 \phi^*$.
With the notation $(AB)_1$ we refer to the contraction of the fields $A$ and $B$ into an $SU(2)_F$ singlet.

The two flavon fields acquire the following vacuum expectation values (VEVs):
\begin{equation} \label{eq:vevs}
    \langle \phi \rangle =
    \begin{pmatrix}
        \varepsilon_{\phi} \Lambda \\
        0 \\
    \end{pmatrix} \hspace{1.3cm} \text{and} \hspace{1.3cm}
    \langle \chi \rangle = \varepsilon_{\chi} \Lambda \ ,
\end{equation}
where we assume that $\varepsilon_{\phi,\chi} \ll 1$ (and they will be used as our expansion parameters) and $\Lambda$ is the cut-off scale of the effective theory. These flavon VEVs determine the presence of the powers of $\varepsilon_{\phi}$ and $\varepsilon_{\chi}$ in the entries of the Yukawa matrices, which bring out the hierarchies.

It was shown in Refs.~\cite{Linster:2018avp,Falkowski:2015zwa} that the quark and charged lepton mass ratios, along with the CKM matrix elements, can be accommodated within their $1\sigma$ experimental ranges at a scale of 10 TeV by taking
\begin{equation} \label{eq:para}
    \varepsilon_{\phi} \sim \lambda^2
    \hspace{1.5cm} \text{and}
    \hspace{1.5cm}
    \varepsilon_{\chi} \sim \lambda^{2 \text{\textdiv} 3} \ ,
\end{equation}
where $\lambda = 0.2$ is roughly the Wolfenstein parameter~\cite{Liu:2014gla}, which correspond, to a first approximation, to the sine of the Cabibbo angle~\cite{Roy:2014nua}. 

Since we are going to use it later on this paper, it is worth showing the Yukawa matrix of the charged lepton sector, obtained from the Lagrangian of eq.~(\ref{eq:lage}):
\begin{equation} \label{eq:yuke}
    Y_e \approx
    \begin{pmatrix}
        e_{11} \varepsilon_{\phi}^2 \varepsilon_{\chi}^4 &
        e_{12} \varepsilon_{\chi}^2 &
        e_{13} \varepsilon_{\phi} \varepsilon_{\chi}^2 \\
        -e_{12} \varepsilon_{\chi}^2 &
        e_{22} \varepsilon_{\phi}^2 &
        e_{23} \varepsilon_{\phi} \\
        e_{31} \varepsilon_{\phi} \varepsilon_{\chi}^3 &
        e_{32} \varepsilon_{\phi} \varepsilon_{\chi} &
        e_{33} \varepsilon_{\chi} \\
    \end{pmatrix} \ ,
\end{equation}
where $e_{ij}$ are (in general) complex $\mathcal{O} (1)$ coefficients.

\section{The neutrino sector revised} \label{sec:neutrinosector}

In this Section, we will assume both the existence of the right-handed (RH) neutrinos and Lepton Number Violation (LNV) interactions.
The three generations of RH neutrinos $N_i$ transform as singlets under the Standard Model gauge group and, in general, as a given representation $N$ under the flavor symmetry group.
In~\cite{Linster:2018avp}, an attempt to apply the $U(2)_F$ to neutrinos was already done. However, they restricted the analysis only on positive (or null) $U(1)_F$ charges and considered only the possibility in which the first two generations of $N$ transform as an $SU(2)_F$ doublet, while the third as a singlet. This latter assumption, which seems to be the most natural (it follows that of left-handed neutrinos, see Table~\ref{tab:quantnum}), lead to predictions that are incompatible with the physical data.
In addition to the previous attempts, we consider here all other possible $SU(2)_F$ and $U(1)_F$ quantum numbers for $N_1$, $N_2$ and $N_3$, on which we impose no constraints related to their signs.
As we will see, this arbitrariness leads to a large number of neutrino mass matrix structures, which we will call \textit{patterns}. Each mass matrix pattern will be analyzed in order to see whether they are able to reproduce neutrino masses and mixing. 

\subsection{Catalog of possible mass matrices} \label{sec:catalog}

\begin{figure} [tb]
    \centering
    \includegraphics[width=0.7\linewidth]{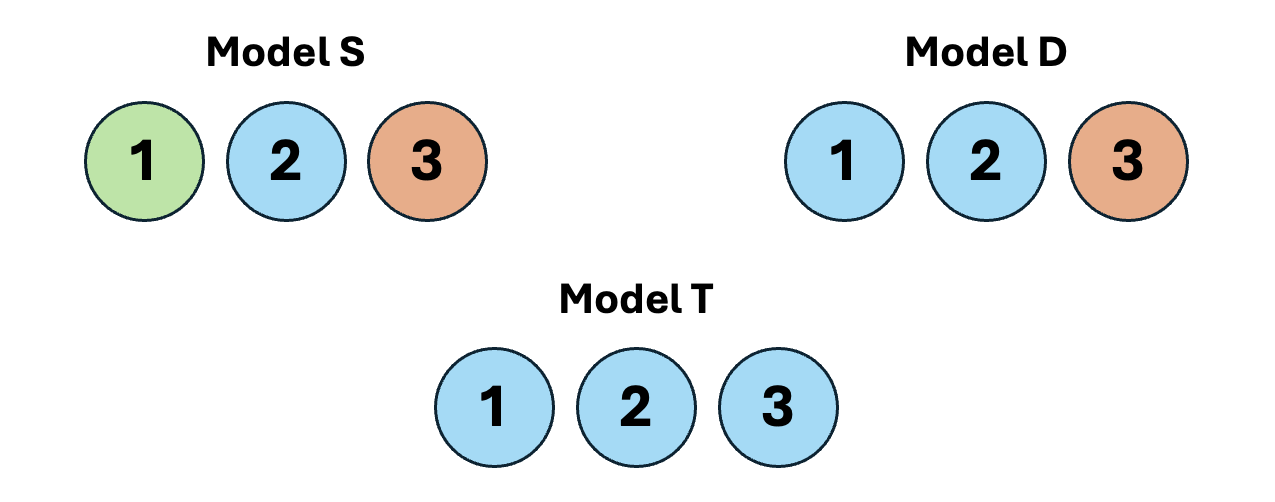}
    \caption{\label{fig:mod} Graphic representation of the three Models that come out in the $SU(2)_F$ application to Majorana neutrinos. The numbers $1,2,3$ correspond to $N_1,N_2,N_3$. Different colours correspond to different $SU(2)_F$ representations (for example, Model \textbf{D} has the blue corresponding to the doublet and the red corresponding to the singlet).}
\end{figure}

The three generations of the right-handed neutrinos can transform under $SU(2)_F$ in three different ways, which correspond to three different Models, as follows:
\begin{itemize}
    \item \textbf{Model \textbf{S}}: $N_1$, $N_2$ and $N_3$ transform as 3 independent $SU(2)_F$ singlets;
    \item \textbf{Model \textbf{D}}: $N_a$ ($a = 1, 2$) transform as a doublet of $SU(2)_F$, while $N_3$ as a singlet;
    \item \textbf{Model \textbf{T}}: $N_i$ ($i = 1,2,3$) transform as a triplet of $SU(2)_F$.
\end{itemize}
A graphical representation of the three Models is reported in Figure~\ref{fig:mod}.
Notice that in Model \textbf{D} one should consider three different sub-cases, corresponding to the three different dispositions of the singlet (it can be $N_1$, $N_2$ or $N_3$). However, it can be demonstrated that these three cases are equivalent and give the same results, so we will consider only the case in which $N_3$ is the singlet, as quoted above.

The Yukawa couplings and the Majorana mass term reads:
\begin{equation} \label{eq:lagnu}
    \mathcal{L}_{\nu} = L^T Y_{\nu} NH + \frac{1}{2} N^T M_{\nu} N + h.c. \ ,
\end{equation}
where $Y_{\nu}$ is the Yukawa matrix, $M_{\nu}$ is the Majorana mass matrix.
In the Model \textbf{S}, all the three generations of $N$ are $SU(2)_F$ singlets, so their $U(1)_F$ charges will be indicated as three different variables $X_1$, $X_2$ and $X_3$.
The $U(2)_F$ invariant Yukawa coupling terms for the Model \textbf{S} are:\footnote{The presence of moduli in this spurion analysis arises from the absence of any assumptions regarding the values of the charges $X_i$.}
\begin{equation} \label{eq:lagSdir}
\begin{split}
    \mathcal{L}_{\nu}^{yuk (S)} &= \frac{y_{11}}{\Lambda^{1+\lvert 2 + X_1 \rvert}} \chi^{\lvert 2 + X_1 \rvert} (\widetilde{\phi} L)_1 N_1 H +
    \frac{y_{12}}{\Lambda^{1+\lvert 2 + X_2 \rvert}} \chi^{\lvert 2 + X_2 \rvert} (\widetilde{\phi} L)_1 N_2 H \\
    &+ \frac{y_{13}}{\Lambda^{1+\lvert 2 + X_3 \rvert}} \chi^{\lvert 2 + X_3 \rvert} (\widetilde{\phi} L)_1 N_3 H +
    \frac{y_{21}}{\Lambda^{1+\lvert X_1 \rvert}} \chi^{\lvert X_1 \rvert} (\phi L)_1 N_1 H \\
    &+ \frac{y_{22}}{\Lambda^{1+\lvert X_2 \rvert}} \chi^{\lvert X_2 \rvert} (\phi L)_1 N_2 H +
    \frac{y_{23}}{\Lambda^{1+\lvert X_3 \rvert}} \chi^{\lvert X_3 \rvert} (\phi L)_1 N_3 H \\
    &+ \frac{y_{31}}{\Lambda^{\lvert 1 + X_1 \rvert}} \chi^{\lvert 1 + X_1 \rvert} L_3 N_1 H +
    \frac{y_{32}}{\Lambda^{\lvert 1 + X_2 \rvert}} \chi^{\lvert 1 + X_2 \rvert} L_3 N_2 H \\
    &+ \frac{y_{33}}{\Lambda^{\lvert 1 + X_3 \rvert}} \chi^{\lvert 1 + X_3 \rvert} L_3 N_3 H \ .    
\end{split}
\end{equation}
For each possible coupling of the Lagrangian, we have shown  the leading order (LO) term only, \textit{i.e.} the one with the least number of flavon fields.  $y_{ij}$ are $\mathcal{O} (1)$ complex coefficients.
For the Majorana mass term we have instead:
\begin{equation} \label{eq:lagSmaj}
\begin{split}
    \mathcal{L}_{\nu}^{M (S)} &= \frac{k_{11}}{\Lambda^{\lvert 2 X_1 \rvert}} \chi^{\lvert 2 X_1 \rvert} N_1 N_1 +
    \frac{k_{12}}{\Lambda^{\lvert X_1 + X_2 \rvert}} \chi^{\lvert X_1 + X_2 \rvert} N_1 N_2 \\
    &+ \frac{k_{13}}{\Lambda^{\lvert X_1 + X_3 \rvert}} \chi^{\lvert X_1 + X_3 \rvert} N_1 N_3 +
    \frac{k_{22}}{\Lambda^{\lvert 2 X_2 \rvert}} \chi^{\lvert 2 X_2 \rvert} N_2 N_2 \\
    &+ \frac{k_{23}}{\Lambda^{\lvert X_2 + X_3 \rvert}} \chi^{\lvert X_2 + X_3 \rvert} N_2 N_3 +
    \frac{k_{33}}{\Lambda^{\lvert 2 X_3 \rvert}} \chi^{\lvert 2 X_3 \rvert} N_3 N_3 \ ,    
\end{split}   
\end{equation}
where $k_{ij}$ are $\mathcal{O} (1)$ complex coefficients.

After inserting the flavon VEVs of eq.~(\ref{eq:vevs}), the dependence on $\Lambda$ disappears and, writing the Lagrangian in the compact form of eq.~(\ref{eq:lagnu}), we derive the following Yukawa matrix and Majorana mass matrix:
\begin{align}
    Y_{\nu}^S =
    \begin{pmatrix} \label{eq:matyukS}
        y_{11} \varepsilon_{\phi} \varepsilon_{\chi}^{\lvert 2 + X_1 \rvert} & y_{12} \varepsilon_{\phi} \varepsilon_{\chi}^{\lvert 2 + X_2 \rvert} &
        y_{13} \varepsilon_{\phi} \varepsilon_{\chi}^{\lvert 2 + X_3 \rvert} \\
        y_{21} \varepsilon_{\phi} \varepsilon_{\chi}^{\lvert X_1 \rvert} &
        y_{22} \varepsilon_{\phi} \varepsilon_{\chi}^{\lvert X_2 \rvert} &
        y_{23} \varepsilon_{\phi} \varepsilon_{\chi}^{\lvert X_3 \rvert} \\
        y_{31} \varepsilon_{\chi}^{\lvert 1 + X_1 \rvert} &
        y_{32} \varepsilon_{\chi}^{\lvert 1 + X_2 \rvert} &
        y_{33} \varepsilon_{\chi}^{\lvert 1 + X_3 \rvert} \\
    \end{pmatrix} \\
    \intertext{and}
    M_{\nu}^S = M
    \begin{pmatrix}
        k_{11} \varepsilon_{\chi}^{\lvert 2 X_1 \rvert} &
        k_{12} \varepsilon_{\chi}^{\lvert X_1 + X_2 \rvert} &
        k_{13} \varepsilon_{\chi}^{\lvert X_1 + X_3 \rvert} \\
        k_{12} \varepsilon_{\chi}^{\lvert X_1 + X_2 \rvert} &
        k_{22} \varepsilon_{\chi}^{\lvert 2 X_2 \rvert} &
        k_{23} \varepsilon_{\chi}^{\lvert X_2 + X_3 \rvert} \\
        k_{13} \varepsilon_{\chi}^{\lvert X_1 + X_3 \rvert} &
        k_{23} \varepsilon_{\chi}^{\lvert X_2 + X_3 \rvert} &
        k_{33} \varepsilon_{\chi}^{\lvert 2 X_3 \rvert} \\
    \end{pmatrix} \ , \label{eq:matmajS}
\end{align}
where $M$ is the mass scale of the RH neutrinos as in the usual see-saw.
It is worth noting that the model is not supersymmetric, so whenever an exponent in the two matrices becomes negative, we can still replace the scalar field $\chi$ with its conjugated to ensure that only positive powers in the expansion parameter $\varepsilon_\chi$ appear.

The light neutrino mass matrix $m_{\nu}^M$ is obtained via the type-I see-saw mechanism~\cite{Brdar:2019iem}, according to the master formula:
\begin{equation} \label{eq:lightmass}
    m_{\nu}^M = - v^2 \ Y_{\nu} M_{\nu}^{-1} Y_{\nu}^T \ .
\end{equation}
As this expression holds for all three models, we omit the superscripts from $Y_{\nu}$ and $M_{\nu}$ to simplify the notation.

The choice of charges for the right-handed neutrinos leads to differences in the parametric structure of the neutrino mass matrix because of the moduli in~(\ref{eq:matyukS}) and~(\ref{eq:matmajS}), which act as a sign change on the arguments of the exponentials when they become negative.
For example, for $X_1,X_2,X_3 \geq 0$ all the quantities within the moduli in $Y_{\nu}^S$ and $M_{\nu}^S$ are $\geq 0$, for $(X_1,X_2,X_3) = (-1,0,0)$ some are negative and for $X_1,X_2,X_3 \le -2$ they are all negative. For each case we will have a different resulting matrix. 
Thus, for the same Model, we get a finite number of different mass matrices, that we call \textit{patterns}. We will discuss more about that in the following.
\vspace{0.3cm}

In Model \textbf{D}, the $U(1)_F$ charges are $X_D$ and $X_3$, associated with the $N_a$ ($a=1,2$) doublet and $N_3$, respectively.
The two terms of the Lagrangian are in this case:
\begin{align}
\begin{split} \label{eq:lagDdir}
    \mathcal{L}_{\nu}^{yuk (D)} &= \frac{y_{11}}{\Lambda^{2+\lvert 3 + X_D \rvert}} \chi^{\lvert 3 + X_D \rvert} (\widetilde{\phi} L)_1 (\widetilde{\phi} N)_1 H +
    \frac{y_{12}}{\Lambda^{\lvert 1 + X_D \rvert}} \chi^{\lvert 1 + X_D \rvert} (L N)_1 H \\
    &+ \frac{y_{13}}{\Lambda^{1+\lvert 2 + X_3 \rvert}} \chi^{\lvert 2 + X_3 \rvert} (\widetilde{\phi} L)_1 N_3 H +
    \frac{y_{22}}{\Lambda^{2+\lvert X_D - 1 \rvert}} \chi^{\lvert X_D - 1 \rvert} (\phi L)_1 (\phi N)_1 H \\
    &+ \frac{y_{23}}{\Lambda^{1+\lvert X_3 \rvert}} \chi^{\lvert X_3 \rvert} (\phi L)_1 N_3 H +
    \frac{y_{31}}{\Lambda^{1+\lvert 2 + X_D \rvert}} \chi^{\lvert 2 + X_D \rvert} L_3 (\widetilde{\phi} N)_1 H \\
    &+ \frac{y_{32}}{\Lambda^{1+\lvert X_D \rvert}} \chi^{\lvert X_D \rvert} L_3 (\phi N)_1 H +
    \frac{y_{33}}{\Lambda^{\lvert 1 + X_3 \rvert}} \chi^{\lvert 1 + X_3 \rvert} L_3 N_3 H    
\end{split} \\
\intertext{and}
\begin{split} \label{eq:lagDmaj}
    \mathcal{L}_{\nu}^{M (D)} &= \frac{k_{11}}{\Lambda^{2+\lvert 2 + 2 X_D \rvert}} \chi^{\lvert 2 + 2 X_D \rvert} (\widetilde{\phi} N)_1 (\widetilde{\phi} N)_1 +
    \frac{k_{12}}{\Lambda^{2+\lvert 2 X_D \rvert}} \chi^{\lvert 2 X_D \rvert} (\widetilde{\phi} N)_1 (\phi N)_1 \\
    &+ \frac{k_{13}}{\Lambda^{1+\lvert 1 + X_D + X_3 \rvert}} \chi^{\lvert 1 + X_D + X_3 \rvert} (\widetilde{\phi} N)_1 N_3 \\
    &+ \frac{k_{22}}{\Lambda^{2+\lvert 2 X_D - 2 \rvert}} \chi^{\lvert 2 X_D - 2 \rvert} (\phi N)_1 (\phi N)_1 \\
    &+ \frac{k_{23}}{\Lambda^{1+\lvert X_D + X_3 - 1 \rvert}} \chi^{\lvert X_D + X_3 - 1 \rvert} (\phi N)_1 N_3 +
    \frac{k_{33}}{\Lambda^{\lvert 2 X_3 \rvert}} \chi^{\lvert 2 X_3 \rvert} N_3 N_3 \ .   
\end{split}    
\end{align}
Inserting the flavon VEVs in eqs.~(\ref{eq:lagDdir}) and~(\ref{eq:lagDmaj}), we obtain
\begin{equation}
    Y_{\nu}^D \approx
    \begin{pmatrix}
        y_{11} \varepsilon_{\phi}^2 \varepsilon_{\chi}^{\lvert 3 + X_D \rvert} & y_{12} \varepsilon_{\chi}^{\lvert 1 + X_D \rvert} &
        y_{13} \varepsilon_{\phi} \varepsilon_{\chi}^{\lvert 2 + X_3 \rvert} \\
        -y_{12} \varepsilon_{\chi}^{\lvert 1 + X_D \rvert} &
        y_{22} \varepsilon_{\phi}^2 \varepsilon_{\chi}^{\lvert X_D - 1 \rvert} &
        y_{23} \varepsilon_{\phi} \varepsilon_{\chi}^{\lvert X_3 \rvert} \\
        y_{31} \varepsilon_{\phi} \varepsilon_{\chi}^{\lvert 2 + X_D \rvert} &
        y_{32} \varepsilon_{\phi} \varepsilon_{\chi}^{\lvert X_D \rvert} &
        y_{33} \varepsilon_{\chi}^{\lvert 1 + X_3 \rvert} \\
    \end{pmatrix}\,,
\end{equation}
and
\begin{equation}
    M_{\nu}^D = M
    \begin{pmatrix}
        k_{11} \varepsilon_{\phi}^2 \varepsilon_{\chi}^{\lvert 2 + 2 X_D \rvert} &
        k_{12} \varepsilon_{\phi}^2 \varepsilon_{\chi}^{\lvert 2 X_D \rvert} &
        k_{13} \varepsilon_{\phi} \varepsilon_{\chi}^{\lvert 1 + X_D + X_3 \rvert} \\
        k_{12} \varepsilon_{\phi}^2 \varepsilon_{\chi}^{\lvert 2 X_D \rvert} &
        k_{22} \varepsilon_{\phi}^2 \varepsilon_{\chi}^{\lvert 2 X_D - 2 \rvert} &
        k_{23} \varepsilon_{\phi} \varepsilon_{\chi}^{\lvert X_D + X_3 - 1 \rvert} \\
        k_{13} \varepsilon_{\phi} \varepsilon_{\chi}^{\lvert 1 + X_D + X_3 \rvert} &
        k_{23} \varepsilon_{\phi} \varepsilon_{\chi}^{\lvert X_D + X_3 - 1 \rvert} &
        k_{33} \varepsilon_{\chi}^{\lvert 2 X_3 \rvert} \\
    \end{pmatrix} \, .
\end{equation}
Note that in the (12) entry of the Majorana mass matrix, the term without $\epsilon_{\phi}^2$ (corresponding to $(N N)_1$) vanishes because of the antisymmetric nature of the $SU(2)_F$ singlet contraction. For this reason, a double insertion of the flavon $\phi$ is required.
The light neutrino mass is given by eq.~(\ref{eq:lightmass}) and, also in this case, there are a finite number of different patterns.
\vspace{0.3cm}

Finally, for Model \textbf{T}, the three RH neutrinos transform as a triplet $N_i$ ($i=1,2,3$) under the $U(1)_F$ symmetry group, with charge $X_T$.
The two $U(2)_F$ invariant terms of the Lagrangian are therefore
\begin{align}
\begin{split} \label{eq:lagTdir}
    \mathcal{L}_{\nu}^{yuk (T)} &= \frac{y_{11}}{\Lambda^{3+\lvert 4 + X_T \rvert}} \chi^{\lvert 4 + X_T \rvert} (\widetilde{\phi} \widetilde{\phi} \widetilde{\phi} L N)_1 H +
    \frac{y_{12}}{\Lambda^{1+\lvert 2 + X_T \rvert}} \chi^{\lvert 2 + X_T \rvert} (\widetilde{\phi} L N)_1 H \\
    &+ \frac{y_{13}}{\Lambda^{1+\lvert X_T \rvert}} \chi^{\lvert X_T \rvert} (\phi L N)_1 H +
    \frac{y_{23}}{\Lambda^{3+\lvert X_T - 2 \rvert}} \chi^{\lvert X_T - 2 \rvert} (\phi \phi \phi L N)_1 H \\
    &+ \frac{y_{31}}{\Lambda^{2+\lvert 3 + X_T \rvert}} \chi^{\lvert 3 + X_T \rvert} L_3 (\widetilde{\phi} \widetilde{\phi} N)_1 H +
    \frac{y_{32}}{\Lambda^{2+\lvert 1 + X_T \rvert}} \chi^{\lvert 1 + X_T \rvert} L_3 (\phi \widetilde{\phi} N)_1 H \\
    &+ \frac{y_{33}}{\Lambda^{2+\lvert X_T - 1 \rvert}} \chi^{\lvert X_T - 1 \rvert} L_3 (\phi \phi N)_1 H
\end{split} \\
\intertext{and}
\begin{split} \label{eq:lagTmaj}
    \mathcal{L}_{\nu}^{M (T)} &= \frac{k_{11}}{\Lambda^{4+\lvert 4 + 2 X_T \rvert}} \chi^{\lvert 4 + 2 X_T \rvert} (\widetilde{\phi} \widetilde{\phi} N \widetilde{\phi} \widetilde{\phi} N)_1 +
    \frac{k_{12}}{\Lambda^{4+\lvert 2 + 2 X_T \rvert}} \chi^{\lvert 2 + 2 X_T \rvert} (\widetilde{\phi} \widetilde{\phi} N \widetilde{\phi} \phi N)_1 \\
    &+ \frac{k_{13}}{\Lambda^{\lvert 2 X_T \rvert}} \chi^{\lvert 2 X_T \rvert} (N N)_1 +
    \frac{k_{23}}{\Lambda^{4+\lvert 2 X_T - 2 \rvert}} \chi^{\lvert 2 X_T - 2 \rvert} (\phi \phi N \phi \widetilde{\phi} N)_1 \\
    &+ \frac{k_{33}}{\Lambda^{4+\lvert 2 X_T - 4 \rvert}} \chi^{\lvert 2 X_T - 4 \rvert} (\phi \phi N \phi \phi N)_1 \ ,
\end{split}    
\end{align}
where the subscript "1" indicates, as before, the singlet contraction of all fields within the parentheses.

If we insert the flavon VEVs in eqs.~(\ref{eq:lagTdir}) and~(\ref{eq:lagTmaj}), we obtain
\begin{equation}
    Y_{\nu}^T \approx
    \begin{pmatrix}
        y_{11} \varepsilon_{\phi}^3 \varepsilon_{\chi}^{\lvert 4 + X_T \rvert} & y_{12} \varepsilon_{\phi} \varepsilon_{\chi}^{\lvert 2 + X_T \rvert} &
        \sqrt{2} y_{13} \varepsilon_{\phi} \varepsilon_{\chi}^{\lvert X_T \rvert} \\
        -\sqrt{2} y_{12} \varepsilon_{\phi} \varepsilon_{\chi}^{\lvert 2 + X_T \rvert} &
        - y_{13} \varepsilon_{\phi} \varepsilon_{\chi}^{\lvert X_T \rvert} &
        y_{23} \varepsilon_{\phi}^3 \varepsilon_{\chi}^{\lvert X_T - 2 \rvert} \\
        y_{31} \varepsilon_{\phi}^2 \varepsilon_{\chi}^{\lvert 3 + X_T \rvert} &
        y_{32} \varepsilon_{\phi}^2 \varepsilon_{\chi}^{\lvert 1 + X_T \rvert} &
        y_{33} \varepsilon_{\phi}^2 \varepsilon_{\chi}^{\lvert X_T - 1 \rvert} \\
    \end{pmatrix}
\end{equation}
and
\begin{equation}
    M_{\nu}^T = M
    \begin{pmatrix}
        k_{11} \varepsilon_{\phi}^4 \varepsilon_{\chi}^{\lvert 4 + 2 X_T \rvert} &
        k_{12} \varepsilon_{\phi}^4 \varepsilon_{\chi}^{\lvert 2 + 2 X_T \rvert} &
        k_{13} \varepsilon_{\chi}^{\lvert 2 X_T \rvert} \\
        k_{12} \varepsilon_{\phi}^4 \varepsilon_{\chi}^{\lvert 2 + 2 X_T \rvert} &
        - k_{13} \varepsilon_{\chi}^{\lvert 2 X_T \rvert} &
        k_{23} \varepsilon_{\phi}^4 \varepsilon_{\chi}^{\lvert 2 X_T - 2 \rvert} \\
        k_{13} \varepsilon_{\chi}^{\lvert 2 X_T \rvert} &
        k_{23} \varepsilon_{\phi}^4 \varepsilon_{\chi}^{\lvert 2 X_T - 2 \rvert} &
        k_{33} \varepsilon_{\phi}^4 \varepsilon_{\chi}^{\lvert 2 X_T - 4 \rvert} \\
    \end{pmatrix} \ ,
\end{equation}
from which we get the light neutrino mass matrix with the various patterns.

\subsubsection{Classification of the matrices}

When we analyzed the reliability of the $U(2)_F$ model on quarks and charged leptons (see Section~\ref{sec:fitquark}), we said that the fermion mass ratios and the CKM entries can be approximately reproduced following the parameterization of the VEVs given in eq.~(\ref{eq:para}).
Therefore, we consider the following two cases:
\begin{gather} \label{eq:par22}
    \varepsilon_{\phi} = \varepsilon_{\chi} = \lambda^2 \ , \\[0.1cm]
    \varepsilon_{\phi} = \lambda^2 \hspace{0.6cm} \text{and} \hspace{0.6cm} \varepsilon_{\chi} = \lambda^3 \ , \label{eq:par23}
\end{gather}
so that all patterns are characterized by the powers of a single parameter $\lambda$.
We refer to the {\it A-Scenario} as that corresponding to eq.~(\ref{eq:par22}) and to the {\it B-Scenario} as that of eq.~(\ref{eq:par23}).

We have therefore the following situation: there are three different Models (Model \textbf{S}, \textbf{D} and \textbf{T}) arising from the transformation properties of the three generations of $N$ under $SU(2)_F$. For each Model, there is a finite number of different patterns arising from different choices of the $U(1)_F$ charges for the three RH neutrinos $N_i$ and these patterns change depending on whether A-Scenario or B-Scenario is considered.

It is worth to notice that there are many cases in which different $U(1)_F$ charge assignments lead to the same mass matrix at LO. 
For example, if we consider Model \textbf{D} with A-Scenario, we have the same LO pattern for different charge assignments $(X_D,X_3) = \{(-2,-1), (-1,-1), (-2,-2)\}$. We have chosen to identify as equal two different mass matrices if they differ by terms that are at least a factor $\lambda^4$ smaller than the leading term.
Taking this into account, the total number of patterns, considering all the possible $U(1)_F$ charges and the two possible Scenarios, are so distributed:
\begin{itemize}
    \item 48 for Model \textbf{S};
    \item 46 for Model \textbf{D};
    \item 10 for Model \textbf{T}.
\end{itemize}
As a result, we can identify 104 distinct structures for the Majorana light neutrino mass matrix.

\subsection{Numerical Analysis}
\label{sec:fit}

We test whether each pattern is able to reproduce the six dimensionless low-energy observables in Table~\ref{tab:par} by performing a fit to the parameter set $\{e_{ij},y_{ij},k_{ij}\}$.\footnote{The mass scale $M$ of the right-handed neutrinos can be inferred from the overall mass scale $v^2/M$ in the light neutrino mass matrix and it is not included in the fit, as it can be easily computed by using either $|\Delta m^2_{\text{atm}}|$ or $\Delta m^2_{\text{sol}}$. %recovered through the $|\Delta m^2_{\text{atm}}|$. 
We considered $|\Delta m^2_{\text{atm}}/(10^{-3}\text{eV}^2)|=2.513^{+0.021}_{-0.019}$ ({\bf{NO}}) and $|\Delta m^2_{\text{atm}}/(10^{-3}\text{eV}^2)|=2.484^{+0.020}_{-0.020}$ ({\bf{IO}}) from NuFit 6.0~\cite{Esteban:2024eli}.
In our scans, $M \in [10^{12}, 10^{14}] \ \text{GeV}$.}
We perform a $\chi^2$ analysis based on 
\begin{equation}
\label{chisq}
\chi^2(p_i)=\sum_{j=1}^6 \left(\frac{q_j(p_i)-q_j^{\text{b-f}}}{\sigma_j}\right)^2\,,
\end{equation}
where $q_j^{\text{b-f}}$ and $\sigma_j$ are, respectively, the best-fit values and the $1\sigma$ uncertainties from Table~\ref{tab:par}. On the other hand, $q_j(p_i)$ are the predicted observables obtained from the parameter set $p_i=\{e_{ij},y_{ij},k_{ij}\}$. 
Since in our approach neutrinos are Majorana particles, in our numerical analysis the mixing angles are extracted from the PMNS mixing matrix $U\equiv {U_L^{\ell}}^{\dag} U_L^{\nu}$ parametrized as:
\begin{equation}
\label{pmns}
U =
\begin{pmatrix}
c_{12} c_{13} & s_{12} c_{13} & s_{13}e^{-i\delta _\text{CP}} \\
-s_{12} c_{23} - c_{12} s_{23} s_{13}e^{i\delta _\text{CP}} &
c_{12} c_{23} - s_{12} s_{23} s_{13}e^{i\delta _\text{CP}} & s_{23} c_{13} \\
s_{12} s_{23} - c_{12} c_{23} s_{13}e^{i\delta _\text{CP}} &
-c_{12} s_{23} - s_{12} c_{23} s_{13}e^{i\delta _\text{CP}} & c_{23} c_{13}
\end{pmatrix}
\text{diag}(e^{-i\alpha_1},e^{-i\alpha_2},1)\,,
\end{equation}
where $c_{ij}\equiv \cos\theta_{ij}$, $s_{ij}\equiv \sin\theta_{ij}$ and $(\alpha_1,\alpha_2)$ are the two Majorana phases.

We assume that the source of CP violation originates solely from the neutrino sector. Therefore, in the numerical analysis, we consider the parameters $k_{ij}$ and $y_{ij}$ as complex numbers, with a free phase ranging from $0$ to $2\pi$; on the other hand, the $e_{ij}$ are taken as real parameters.
It is worth noting that the CP violating phase $\delta_\text{CP}$ has not been considered in the fit due to its current large uncertainty~\cite{Esteban:2024eli} which does not exclude, at high significance, any viable value and will only be measured by future oscillation experiments \cite{DUNE:2020jqi,Hyper-Kamiokande:2018ofw,Alekou:2022emd,ESSnuSB:2023ogw,Agarwalla:2022xdo}; thus, the phases in the parameters are not subject to experimental constraints.
To explore the parameter space and minimize the $\chi^2$ function in eq.~(\ref{chisq}), we used an algorithm inspired by the approach presented in Ref.~\cite{Novichkov:2021cgl}. The fitting procedure uses seed values for the model parameters within the range $[\lambda,\lambda^{-1}]$ and then explores the parameter space by mimicking the Brownian-like evolution of a particle moving in a potential shaped by the figure of merit $l(p_i)=\sqrt{\chi^2(p_i)}$, where $p_i$ is the array of free dimensionless parameters of the model. We also required that the Yukawa hierarchy arises solely by the $U(2)_F$ flavor symmetry breaking, \textit{i.e.~}all free parameters should be $\mathcal{O}(1)$. To enforce this, we introduced the \emph{Parameter of Metric Goodness} ($\text{P}_\mathrm{MG}$), which we define as:
\begin{equation}
    \text{P}_{\mathrm{MG}}=\sum_j \text{log}^2(|\gamma_j^{p_i}|)\,,
\end{equation}
where $\gamma_j^{p_i}$ is any parameter belonging to the set $p_i$. 
This measure accounts for the absolute value of each free parameter in the model.
We consider a fit to be satisfactory if there is at least one set of parameters for which $\chi^2\lesssim 20$ and $\text{P}_\mathrm{MG}\lesssim 30$, given the number of parameters in our models.\footnote{We stress that the primary purpose of this parameter is to discard those configurations in which either all parameters exceed the $\mathcal{O}(1)$, or most of the parameters lie within the desired range but at least one of them is extremely large.}
\begin{table}[tb]
    \centering
    \renewcommand{\arraystretch}{1.5}
    \begin{tabular}{l|c c}
    
         \Xhline{1.5pt}
         \textbf{Parameter} & \textbf{bfp $\pm 1\sigma$ NO} & \textbf{bfp $\pm 1\sigma$ IO} \\
         \hline
         $\theta_{12}/^{\circ}$ & $33.68^{+0.73}_{-0.70}$ & $33.68^{+0.73}_{-0.70}$ \\
         %\hline
         $\theta_{23}/^{\circ}$ & $43.3^{+1.0}_{-0.8}$ & $47.9^{+0.7}_{-0.9}$ \\
         %\hline
         $\theta_{13}/^{\circ}$ & $8.56^{+0.11}_{-0.11}$ & $8.59^{+0.11}_{-0.11}$ \\
         %\hline
         $\alpha\equiv \Delta m^2_\text{sol}/|\Delta m^2_{\text{atm}}|$ & $0.0298 \pm 0.0008$ & $0.0302 \pm 0.0008$ \\
         \hline
         $%r_{12}\equiv
         m_e/m_\mu$ & \multicolumn{2}{c}{$0.0048 \pm 0.0002$} \\
         %\hline
         $%r_{23}\equiv
         m_\mu/m_\tau$ & \multicolumn{2}{c}{$0.0565 \pm 0.0045$} \\
         \Xhline{1.5pt}
    \end{tabular}
    \caption{\label{tab:par} Neutrino observables and charged lepton mass ratios with their $1\sigma$ ranges. Neutrino observables are taken from NuFIT 6.0 using the dataset with Super-Kamiokande atmospheric data~\cite{Esteban:2024eli}, while the mass ratios of charged leptons are taken from~\cite{Feruglio:2017spp}. Here $\Delta m^2_\text{sol}\equiv (m_2^2-m_1^2)$ and 
    %$\Delta m^2_\text{atm}\equiv |m_3^2-(m_2^2-m_1^2)/2|$.
    $\Delta m^2_\text{atm}\equiv |m_3^2-m_j^2|$ ($j=1$ for Normal Ordering and $j=2$ for Inverted Ordering).
    Also, the "Normal Ordering" and the "Inverted Ordering" for the neutrino mass hierarchy are indicated with {\bf{NO}} and {\bf{IO}}, respectively. The running of low-energy neutrino observables can be neglected since in our fits we do not obtain a quasi-degenerate spectrum \cite{Antusch:2003kp}.}
    \renewcommand{\arraystretch}{1}
\end{table}

We obtained 13 viable patterns, all of which provide valid results only in Normal Ordering ({\bf{NO}}) hypothesis, 6 for Model \textbf{S} and 7 for Model \textbf{D}; we will analyzed them in detail in the next Subsection, see Table~\ref{tab:goodpat}. No patterns for Model \textbf{T} were capable of reproducing neutrino data.
In the Appendix~\ref{ap:tab}, we report the fit results in Tables~\ref{tab:BFmodS} and~\ref{tab:BFmodD}, from which it can be seen that each of the 13 models yields a very small $\chi^2$ value. This is not unexpected, as the large number of free parameters makes it likely to find configurations that reproduce the observables used as input. However, a $\chi^2<1$ alone does not guarantee the quality of the fit, since even with a large parameter space the fit might still suffer from a high degree of fine-tuning. To keep this aspect under control, we use the normalised Altarelli-Blankenburg measure $d_\text{FT}$~\cite{Altarelli:2010at}:
\begin{equation}
    d_\text{FT}=\frac{\displaystyle\sum_i\left|\frac{\text{par}_i}{\delta\text{par}_i}\right|}{\displaystyle\sum_i\left|\frac{\text{obs}_i}{\sigma_i}\right|}\,.
    \label{dft}
\end{equation}
The fine-tuning measure defined in eq.~(\ref{dft}) quantifies the sensitivity of the fit to variations in the model parameters. Specifically, for each parameter, $|\text{par}_i|$ denotes its best-fit value, while $|\delta \text{par}_i|$ represents the amount by which the parameter must be shifted—keeping all others fixed—to increase the $\chi^2$ by one unit above its minimum. If even a small deviation from the best-fit value results in a significant increase in $\chi^2$, the model is considered fine-tuned. To make this measure comparable across different scenarios, it is normalised by the sum of the ratios between the best-fit values of the observables and their associated uncertainties.

Comparing Tables~\ref{tab:BFmodS} and~\ref{tab:BFmodD}, we find that the patterns of Model \textbf{D} are more fine-tuned that those of Model \textbf{S}, because the latter have a lower $d_\text{FT}$. In particular, this parameter is very large for the patterns \textbf{D}3 A and \textbf{D}5 B, about which we will discuss in the next Section.

It is worth to notice that all the values of $\chi^2 + \text{P}_\mathrm{MG}$ reported on the last row of the tables in Appendix~\ref{ap:tab} are less than 30, so we can state that the Yukawa hierarchies are mainly due to the $U(2)_F$ breaking.

%initial hypothesis of all coefficients being $\mathcal{O}(1)$ has been satisfied.

\subsection{Viable patterns}

As we mentioned above, our numerical fit procedure returned 13 viable patterns in the hypothesis of {\bf{NO}}. This is the first prediction of the $U(2)_F$ model of flavor: the inverted ordering (IO) scenario is not compatible with the physical oscillation parameters and the charged lepton mass ratios.
The structure of the 13 viable patterns, in which each entry of the matrix contains the leading order (LO) term in $\lambda$, are reported in Table~\ref{tab:goodpat}, where we have grouped the matrices with the same structure in terms of the parameter $\lambda$. We assigned a nomenclature to classify the different patterns: the first letter in the entries of the first column indicates the transformation properties of $(N_1,N_2,N_3)$ under $SU(2)_F$, the consecutive number differentiates between the various patterns for the same Model and the final letter corresponds to the Scenario (A for eq.~(\ref{eq:par22}) and B for eq.~(\ref{eq:par23})).
\renewcommand{\arraystretch}{1}
\begin{table}[t]
    \centering
    \begin{tabular}{l|c c}
    
        \Xhline{1.5pt} \rule[-0.2cm]{0pt}{0.7cm}
        \textbf{Pattern} & \textbf{Charges} & \textbf{LO mass matrix in terms of $\lambda$} \\
        
        \hline \rule[-1.15cm]{0pt}{2.5cm}
        \makecell[c]{\textbf{S}1 A \\ \textbf{S}2 A \\ \textbf{D}1 A \\ \textbf{D}2 A} &
        \makecell[c]{$(1,0,-2)$ \\ $(1,1,-2)$ \\ $(1,-2)$ \\ $(2,-2)$}
        &
        $\begin{pmatrix}
            \lambda^4 & \lambda^4 & \lambda^4 \\
            \lambda^4 & \lambda^4 & \lambda^4 \\
            \lambda^4 & \lambda^4 & \lambda^4 \\
        \end{pmatrix}$
        \\
        
        \hline \rule[-0.85cm]{0pt}{2cm}
        \makecell[c]{\textbf{S}3 A \\ \textbf{S}4 A}
        &
        \makecell[c]{$(2,1,-2)$ \\ $(2,2,-2)$}
        & 
        $\begin{pmatrix}
            \lambda^{12} & \lambda^8 & \lambda^8 \\
            \lambda^8 & \lambda^4 & \lambda^4 \\
            \lambda^8 & \lambda^4 & \lambda^4 \\
        \end{pmatrix}$
        \\
        
        \hline \rule[-0.85cm]{0pt}{2cm}
        \makecell[c]{\textbf{D}3 A \\ \textbf{D}4 A}
        &
        \makecell[c]{$(0,1)$ \\ $(1,0)$}
        & 
        $\begin{pmatrix}
            \lambda^4 & 1 & \lambda^4 \\
            1 & 1/\lambda^4 & 1 \\
            \lambda^4 & 1 & \lambda^4 \\
        \end{pmatrix}$
        \\
        
        \hline \rule[-0.85cm]{0pt}{2cm}
        \makecell[c]{\textbf{S}1 B \\ \textbf{S}2 B}
        &
        \makecell[c]{$(1,0,-2)$ \\ $(1,1,-2)$}
        & 
        $\begin{pmatrix}
            \lambda^4 & \lambda^4 & \lambda^5 \\
            \lambda^4 & \lambda^4 & \lambda^5 \\
            \lambda^5 & \lambda^5 & \lambda^6 \\
        \end{pmatrix}$
        \\

        \hline \rule[-0.85cm]{0pt}{2cm}
        \makecell[c]{\textbf{D}1 B \\ \textbf{D}2 B}
        &
        \makecell[c]{$(1,-2)$ \\ $(2,-2)$}
        & 
        $\begin{pmatrix}
            \lambda^8 & \lambda^6 & \lambda^7 \\
            \lambda^6 & \lambda^4 & \lambda^5 \\
            \lambda^7 & \lambda^5 & \lambda^6 \\
        \end{pmatrix}$
        \\

        \hline \rule[-0.85cm]{0pt}{2cm}
        \makecell[c]{\textbf{D}5 B}
        &
        $(0,0)$
        & 
        $\begin{pmatrix}
            \lambda^8 & \lambda^2 & \lambda^7 \\
            \lambda^2 & \lambda^4 & \lambda \\
            \lambda^7 & \lambda & \lambda^6 \\
        \end{pmatrix}$
        \\
        \Xhline{1.5pt}       
    \end{tabular}
    \caption{\label{tab:goodpat} The 13 viable patterns after the numerical fit, with their corresponding $U(1)_F$ charges ($(X_1,X_2,X_3)$ for Model \textbf{S} and $(X_D,X_3)$ for Model \textbf{D}). In the right column, the corresponding mass matrices with the entries at LO are given in terms of the parameter $\lambda$ (we remind that $\lambda = 0.2$).}
\end{table}
\renewcommand{\arraystretch}{1}

The fact that there are no patterns for Model \textbf{T} is not surprising: in Section~\ref{sec:catalog}, we stated that this Model has only one $U(1)_F$ charge $X_T$, so the resulting mass matrices have fewer parameters than those of Model \textbf{S} and \textbf{D} making the fit to the neutrino observables less successful. 

An interesting feature belonging to all good patterns is that the absolute value of the $U(1)_F$ charges of RH neutrinos  does not exceed 2. 
Furthermore, we can see from the first row of Table~\ref{tab:goodpat} that 4 good patterns have an anarchical structure. The anarchical models can be considered as a particular case of degenerate models with $m_i^2 \sim \Delta m_{atm}^2$. In this class of models, mass degeneracy is replaced by the principle that all mass matrices are structureless in the neutrino sector~\cite{Altarelli:2004za}.
The neutrino mass anarchy has been studied in the literature, see \textit{e.g.} \cite{Hall:1999sn,Haba:2000be,Altarelli:2002sg,deGouvea:2003xe,deGouvea:2012ac,Altarelli:2012ia,Babu:2016aro,Fortin:2017iiw}. In many cases, it has been shown that experimental data can be accounted for by anarchical matrices, which can also provide successful predictions \cite{Lu:2014cla,Jeong:2012zj}.

If we perform a rough perturbative analysis of the 13 viable patterns of Table~\ref{tab:goodpat} considering only the predicted values of the solar to atmospheric mass differences ratio $\alpha$\footnote{Due to the large number of parameters involved in the neutrino sector, we will not show the cumbersome expressions for the neutrino observables in our cases.}, it can be verified that the structure of the patterns \textbf{D}3 A, \textbf{D}4 A and \textbf{D}5 B leads to $\alpha$'s in disagreement with the expected value of $\sim 0.03$ (see Table~\ref{tab:par}).
In particular, the \textbf{D}5 B pattern forces $\alpha$ to take values very close to 1, while the structure of \textbf{D}3 A and \textbf{D}4 A leads to $\alpha \sim \lambda^{16}$. The latter two patterns are not considered viable in~\cite{Linster:2018avp}.
However, in our numerical analysis, the patterns \textbf{D}3 A and \textbf{D}4 A have been found to be compatible with experimental observables at the cost of forcing some parameters to be small.
For instance, in \textbf{D}3 A, the largest entry of the neutrino mass matrix is $(m_{\nu}^M)_{22} \propto k_{23}^2/\lambda^4$. Thus, lowering only $\lvert k_{23} \rvert$ and making it close to the minimum value considered acceptable ($k_{23}\sim\lambda$) it is possible to reduce the gap between the corresponding large eigenvalue (which in NO is $m_3$) and the other two obtaining a reasonable $\alpha$.
For the same reason, in \textbf{D}4 A in order to lower the largest entry of the neutrino mass matrix $(m_\nu^M)_{22}$ we need to make the combination $k_{23}^2-k_{22}k_{33}$ as small as possible, while in \textbf{D}5 B we need to decrease $\lvert k_{33} \rvert$ and $\lvert y_{32} \rvert$ to obtain a value of $\alpha$ different from 1. For the patterns \textbf{D}3 A and \textbf{D}5 B, this aspect also manifests in a large value of the fine-tuning parameter $d_{\text{FT}}$, as we previously anticipated. As we will see in the next Subsection, these requirements on the parameters also cause the lowering of the three eigenvalues, and therefore of the lightest mass $m_L$. The best fit value for the PMNS matrix phase $\delta$ is not shown in our results and will not be discussed further since we checked that there is actually no preferred value for it. 

\subsection{Predictions on neutrino observables}

\subsubsection{Predictions on $m_{\beta\beta}$} \label{sec:mbb}

An important role in the study of neutrino physics is played by neutrinoless double beta ($0\nu2\beta$) decays, which would confirm the Majorana nature of these particles.
In the hypothesis of the existence of light Majorana neutrinos, the rate of the $0\nu2\beta$ decay would depend on the so-called \textit{effective Majorana mass parameter} $m_{\beta\beta}$, which depends on neutrino masses, mixing angles and CP-violating phases\footnote{While in neutrino oscillation the only phase involved in the transition probability is the PMNS matrix Dirac phase $\delta$, the effective Majorana mass depends also on two additional phases which we consider in the numerical analyses to be unbounded.
%coming from the Majorana mass matrix.
} in the following way~\cite{Mei:2024kvs}:
\begin{equation}
    \lvert m_{\beta\beta} \rvert = \left| \sum_i (U_{ei})^2 m_i \right| \ ,
\end{equation}
where $U_{ei}$ are the PMNS elements of the first row.

\begin{figure} [ht]
    \centering
    \includegraphics[width=0.75\textwidth]{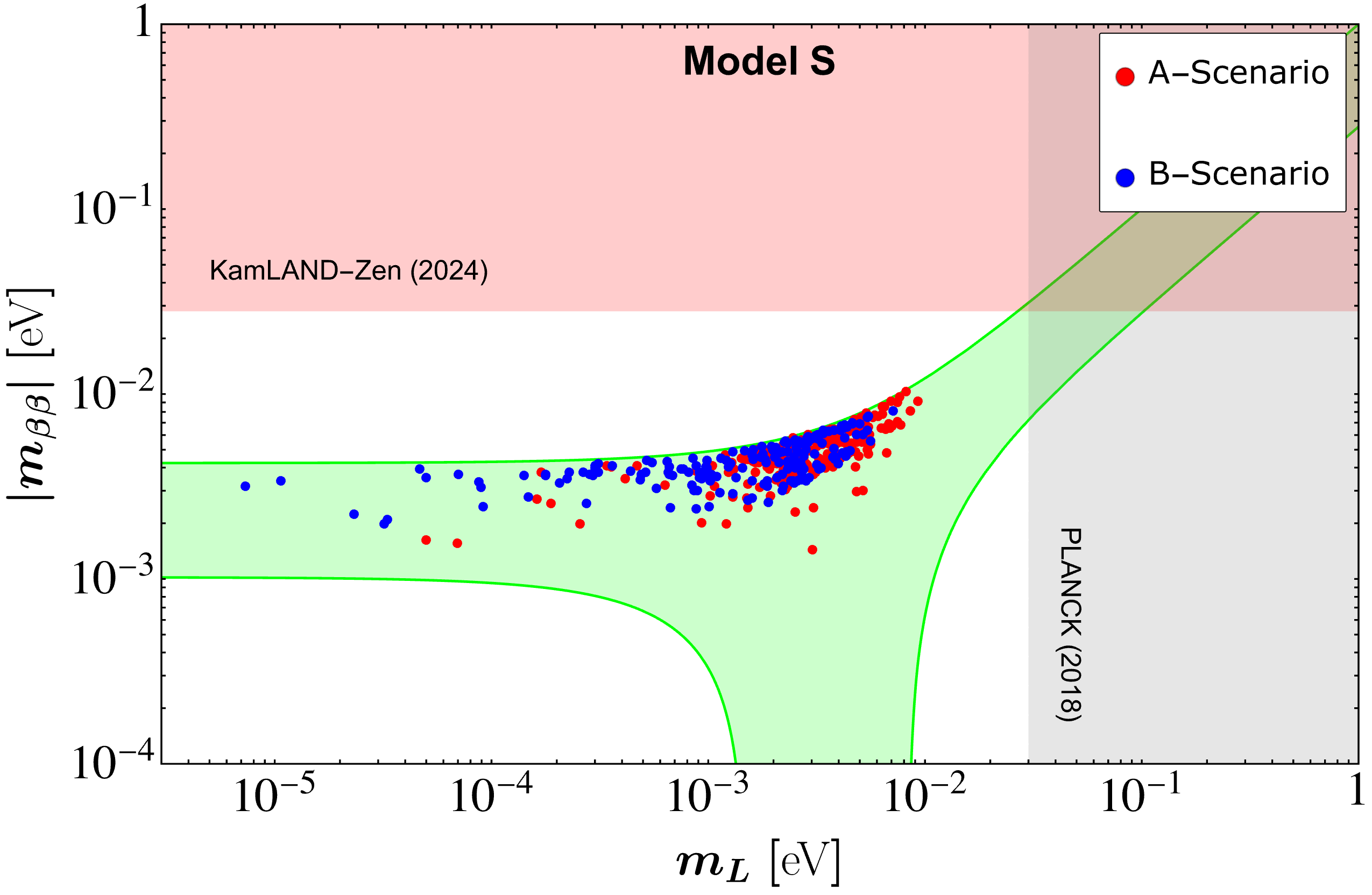} \\[3pt]
    \includegraphics[width=0.75\textwidth]{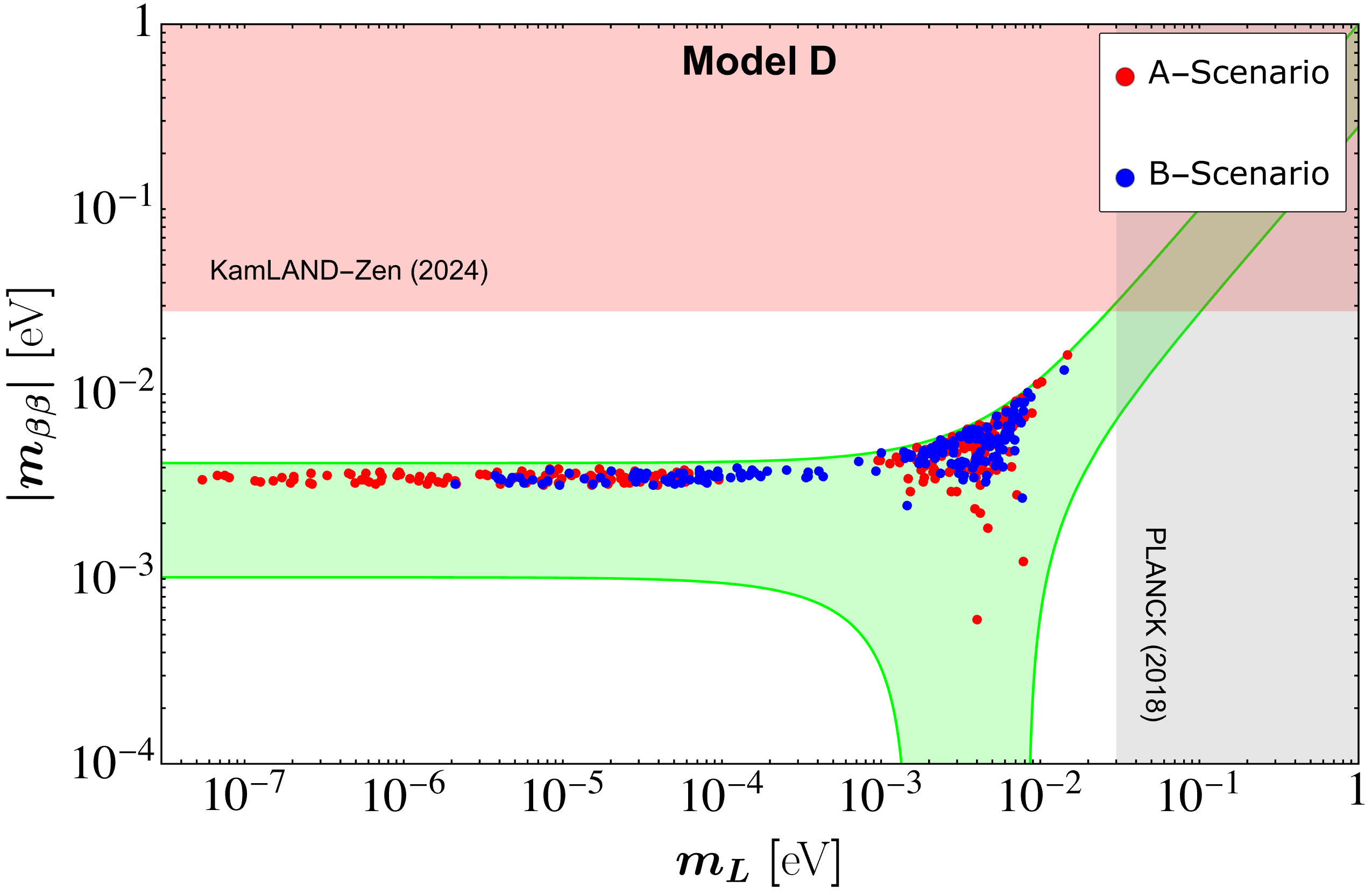}
    \caption{\label{fig:mbbplot} Distributions of the absolute value of $m_{\beta\beta}$ in terms of the lightest neutrino mass $m_L$. For each plot, all the valid representations for the viable patterns belonging to the given Model are collected, with a red or blue point, respectively, for Scenario A or B. The green zone corresponds to the $3\sigma$ confidence region for {\bf{NO}}.
    The red exclusion region corresponds to the latest KamLAND-Zen upper limit on $\lvert m_{\beta\beta} \rvert$~\cite{KamLAND-Zen:2024eml}, while the grey exclusion region corresponds to the PLANCK 2018 upper limit on $m_L$~\cite{Planck:2018vyg,DiValentino:2019dzu}.
    }
\end{figure}

In Fig.~\ref{fig:mbbplot}, we present the plots of $\lvert m_{\beta\beta} \rvert$ versus $m_L$, which is the lightest neutrino mass ($m_1$ for the {\bf{NO}} hypothesis). We classified the 13 viable patterns into four categories based on the Model (\textbf{S} or \textbf{D}) and the Scenario (A or B).
If we call \textit{representation} a particular set of parameters $\{e_{ij}, y_{ij}, k_{ij}\}$, from which we can derive the corresponding mass matrix $m_{\nu}^M$ and PMNS matrix $U$, we define such a representation as valid if it reproduces the six observables of Table~\ref{tab:par} within the experimental $3\sigma$ ranges, which are taken from NuFIT 6.0~\cite{Esteban:2024eli} using the dataset with SK atmospheric data.
Each point in the two plots corresponds to a valid representation; the point is red for A-Scenario (see eq.~(\ref{eq:par22})) or blue for B-Scenario (see eq.~(\ref{eq:par23})).
Each plot gathers the valid representations of all the viable patterns corresponding to the given Model (\textbf{S} or \textbf{D}).
%, because Model \textbf{T} has no viable patterns).
We can note from the plots that there are no points with $\lvert m_{\beta\beta} \rvert > 20 \ \text{meV}$. Currently, the most stringent experimental constraint comes from KamLAND-Zen, which provides an upper bound on $\lvert m_{\beta\beta} \rvert$ of 28 meV at $90 \%$ confidence level (CL)~\cite{KamLAND-Zen:2024eml} (see the red exclusion region in the two plots of Fig.~\ref{fig:mbbplot}). However, upcoming experiments, such as CUPID~\cite{CUPID:2022jlk}, LEGEND-1000~\cite{LEGEND:2021bnm} and nEXO~\cite{nEXO:2021ujk}, aim to reach sensitivities beyond $10 \ \text{meV}$. In particular, the nEXO experiment is designed to reach an upper limit on $\lvert m_{\beta\beta} \rvert$ down to 4.7 meV, which nonetheless is not enough to probe most of the possible $\lvert m_{\beta\beta} \rvert$ obtainable in our models.

In Fig.~\ref{fig:mbbplot} we can also see an upper bound on the lightest mass $m_{L}$ (grey exclusion region), arising from PLANCK 2018 cosmology measurements combined with the Baryonic Acoustic Oscillations (BAO) data~\cite{Planck:2018vyg,DiValentino:2019dzu}, which provide an upper bound $\sum_i m_i \le 0.12 \ \text{eV}$ at $95 \%$ CL. This bound translates to the bound on the lightest neutrino mass, that is, for {\bf{NO}}, $m_L \lesssim 30 \ \text{meV}$. 
These bounds will be improved by the combination of the future experiment results on CMB, galaxy surveys and dark matter measurements. Some examples are given by CMB-S4~\cite{CMB-S4:2016ple,Calabrese:2016eii,Dvorkin:2022bsc}, CORE~\cite{CORE:2016npo}, EUCLID~\cite{Audren:2012vy,Sprenger:2018tdb}, the Large Synoptic Survey Telescope (LSST)~\cite{LSST:2008ijt}, DESI~\cite{Font-Ribera:2013rwa,DESI:2024mwx} and the Square Kilometre Arrey (SKA)~\cite{Zegeye:2024jdc}. These future experiments may be able to probe smaller masses; however, it is difficult to provide a precise estimate of the future upper bounds on $\sum_i m_i$ since these results strongly depend on the cosmological model. An example is given by~\cite{DESI:2025ejh}, where the new DESI measures provide negative neutrino masses as the best-fit values. This is a strong indication that the cosmological model used to derive those values may be incorrect.

If we look at the Model \textbf{D} plot (lower panel in Fig.~\ref{fig:mbbplot}), there is a tail disconnected from the point cluster in both Scenarios. This tail reaches very small values of the lightest mass $m_L$.
For the Model \textbf{D} patterns with A-Scenario (red points), this tail is due to the \textbf{D}3 A and \textbf{D}4 A patterns. As already mentioned, these two patterns have some Majorana parameters forced to take small values, and this constraints lead to very low $m_L$ (some points in the plot reach $m_L < 10^{-7} \ \text{eV}$). We can therefore say that patterns \textbf{D}3 A and \textbf{D}4 A, which in~\cite{Linster:2018avp} are not considered compatible with the observables, are able to reproduce the experimental data at the cost of forcing the value of some Majorana parameters. As a consequence, these patterns predict very small values of the lightest mass $m_L$.
In a similar fashion, it can be found that the tail for the B-Scenario (the blue points in Fig.~\ref{fig:mbbplot}) is due to the pattern \textbf{D}5 B. In this case, however, the $m_L$ values do not fall below $10^{-6} \ \text{eV}$.

It is also interesting to note that, disregarding the tail of Model \textbf{D}, the cluster of points is slightly more dispersed for Model \textbf{S} patterns (upper panel in Fig.~\ref{fig:mbbplot}). This is due to the fact that Model \textbf{S} has two more free parameters than Model \textbf{D}.

The predictions for $m_{\beta\beta}$ in Fig.~\ref{fig:mbbplot} are largely compatible with those of the $D_6 \times U(1)_F$ model with Majorana neutrinos proposed in~\cite{Linster:2018avp}, with the exception of the A-scenario. In that case, some of the predicted values of $m_{\beta\beta}$ fall below the range reported in~\cite{Linster:2018avp}. We can also compare our results with the $U(2)_5$ scenario in~\cite{Linster:2020fww}. The prediction range of the latter extends to slightly lower values than those obtained in our model, especially in the B-scenario. Conversely, regarding the predicted ranges for the sum of the neutrino masses, we find a good level of compatibility with both the model in~\cite{Linster:2018avp} and that in~\cite{Linster:2020fww}.\footnote{Notice that, in the two plots of Fig.~\ref{fig:mbbplot}, only the lightest mass $m_L$ is presented; however, it can be easily converted into the sum of the neutrino masses $\sum m_i$.}

\subsubsection{Predictions on $m_{\beta}$}

\begin{figure} [ht]
    \centering
    \includegraphics[width=0.75\textwidth]{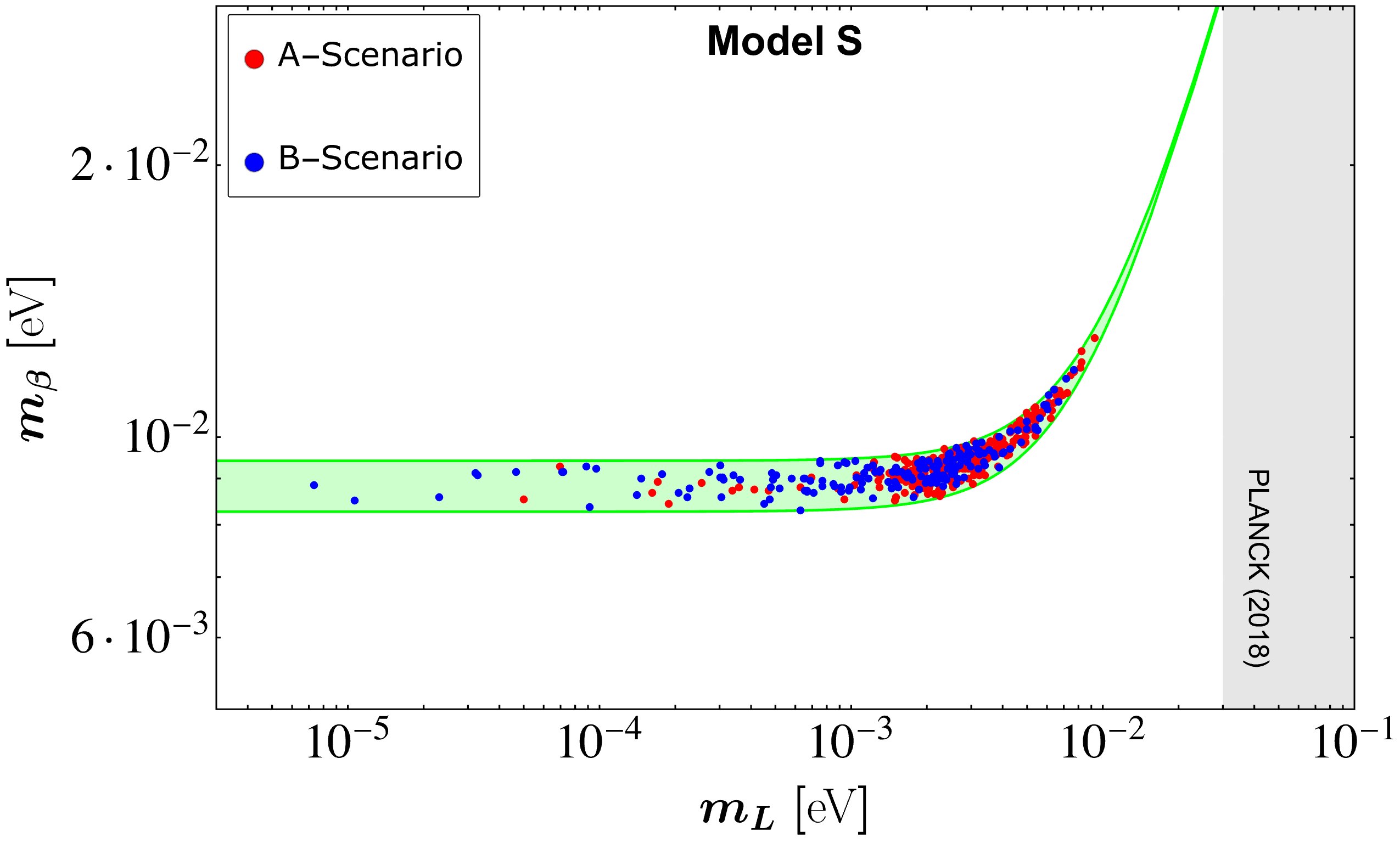} \\[3pt]
    \includegraphics[width=0.75\textwidth]{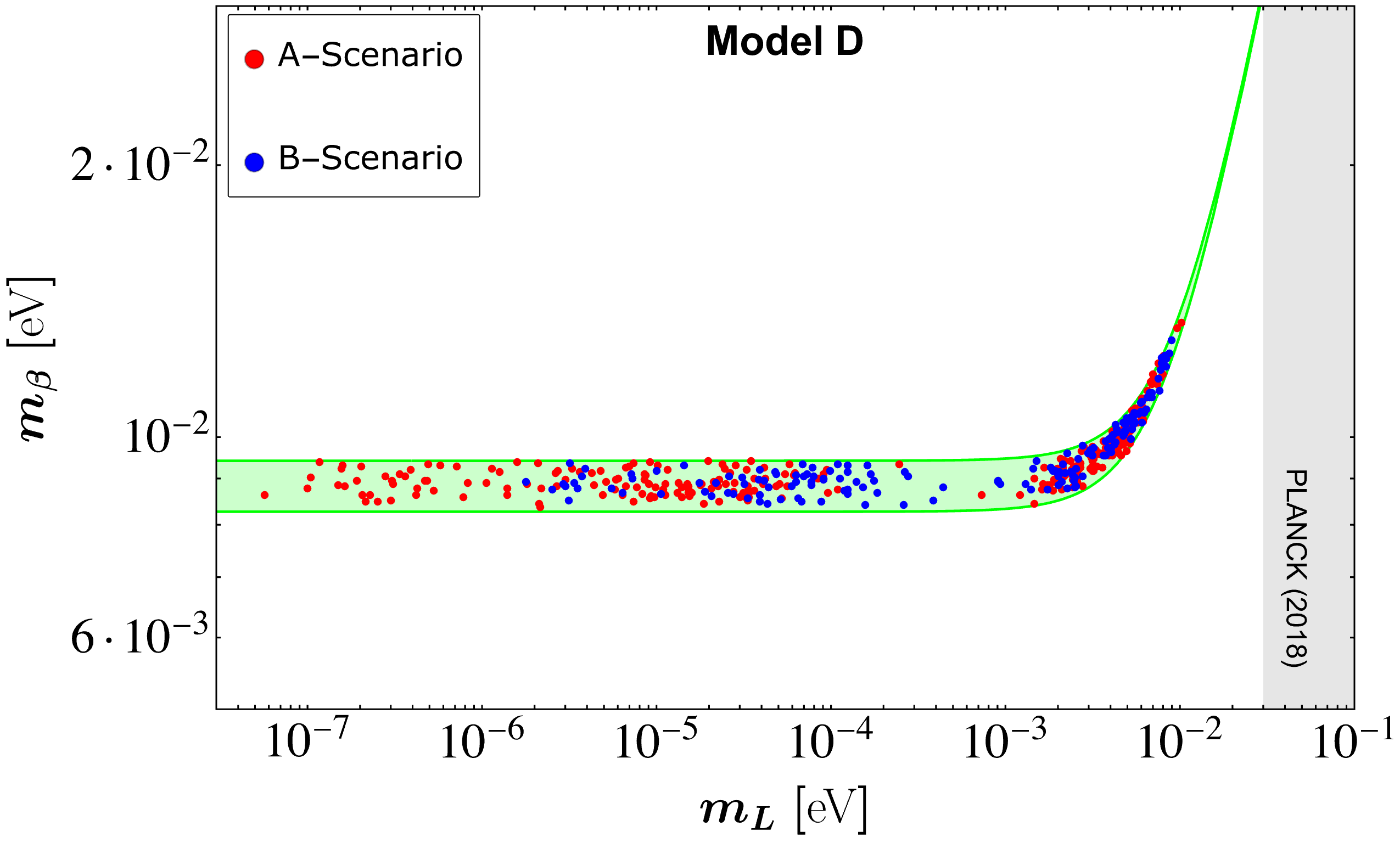}
    \caption{\label{fig:mbplot} Distribution of $m_{\beta}$ in terms of the lightest neutrino mass $m_L$. For each plot, all the valid representations for the viable patterns belonging to the given Model are collected, with a red or blue point, respectively, for Scenario A or B. The green zone corresponds to the $3\sigma$ confidence region for {\bf{NO}}, while the grey exclusion region corresponds to the PLANCK 2018 upper limit on $m_L$~\cite{Planck:2018vyg,DiValentino:2019dzu}.}
\end{figure}

We now turn our interests on the \textit{effective electron neutrino mass} $m_{\beta}$, which determines the endpoint of the beta decay spectrum. This parameter is defined in the following way~\cite{Franklin:2018adt}:
\begin{equation}
    m_{\beta} = \sqrt{\sum_i \lvert U_{ei} \rvert^2 m_i^2} \ .
\end{equation}
The best constraint comes from KATRIN experiment, which investigates the kinematics of the tritium beta spectrum near its kinematical endpoint~\cite{Katrin:2024tvg}. 
In Fig.~\ref{fig:mbplot} we report our results for $m_{\beta}$ versus $m_L$ for both Model \textbf{S} and Model \textbf{D} viable patterns.
Also in this case, each point corresponds to a particular valid representation (red for A-Scenario and blue for B-Scenario). From the two plots, it emerges that all the $m_{\beta}$ predictions are below 20 meV, a value significantly lower than the KATRIN upper bound $m_{\beta} < 450 \ \text{meV}$ (at $90\%$ CL), which is not shown in the plots.
There are some interesting experiments in the pipeline. The most important are the Project 8 experiment~\cite{Project8:2022wqh}, which aims to reach the sensitivity of 40 meV for $m_{\beta}$, and the ECHo~\cite{Gastaldo:2013wha,Mantegazzini:2023igy} and HOLMES experiments~\cite{Alpert:2014lfa,HOLMES:2016spk}, which exploit the electron capture of $^{163}\text{Ho}$ and are expected to reach a sensitivity to $m_{\beta}$ on the order of 100 meV.
The grey exclusion region in the two plots corresponds to the PLANCK 2018 upper bound ($m_L \lesssim 30 \ \text{meV}$), which we have already discussed in Section~\ref{sec:mbb}.

The disconnected tail at very low values of $m_L$ in the Model \textbf{D} plot (lower panel in Fig.~\ref{fig:mbplot}) is caused by the same patterns discussed for the case of $\lvert m_{\beta\beta} \rvert$.

As for $m_{\beta\beta}$, we can likewise compare our range of predictions for $m_{\beta}$ (see the two plots in Fig.~\ref{fig:mbplot}) with the similar models. In this case, together with the $U(2)_5$ model of~\cite{Linster:2020fww} and the $D_6 \times U(1)_F$ model of~\cite{Linster:2018avp}, we can also make the comparison with the $U(2)_F$ model with Dirac neutrinos in~\cite{Linster:2018avp}.
In all of these cases, we find a good level of compatibility between the two prediction ranges.
Regarding the sum of the masses $\sum m_i$, the same conclusions drawn for the plots in Fig.~\ref{fig:mbbplot} hold.

\section{LFV decays and the muon anomalous magnetic moment}\label{LFVintr}
The electric and magnetic dipole moments offer valuable low-energy insights into potential New Physics (NP) beyond the Standard Model. Over the years, the long-standing tension in the anomalous magnetic moment of the muon $a_\mu=(g-2)_\mu/2$ has emerged as one the most compelling probe of NP. After the analyses performed on Run-2 and Run-3 from 2019 to 2020 at Fermilab~\cite{Muong-2:2023cdq}, a new result has been recently published by the collaboration using data up to 2023~\cite{Muong-2:2025xyk}.  Systematic uncertainties were reduced by more than a factor of four compared to the previous world average from the earlier E989 measurements~\cite{Muong-2:2021ojo} and the previous results from BNL~\cite{Muong-2:2006rrc}.

The final result quoted by the Muon $g-2$ collaboration at Fermilab on the muon anomalous magnetic moment~\cite{Muong-2:2025xyk}  is $a_\mu=1165920705(148)\times 10^{-12}$. While this result remains compatible with the previous measurement, new developements in the theoretical calculations within the SM have strongly reduced the tension between data and theory. Indeed, considering the new theory result~\cite{Aliberti:2025beg} $a_\mu^{\text{SM}}=116592033(63)\times 10^{-11}$, the $\Delta a_\mu$ is now vanishing within $1\sigma$,
\begin{equation}
    \Delta a_\mu=a_\mu^\text{Exp}-a_\mu^\text{SM}=(38\pm63)\times 10^{-11}\,.
    \label{eq:g2anomaly}
\end{equation}
It is worth to mention that the previous comparison between SM predictions and the experimental value claimed a 5.1$\sigma$ discrepancy with the SM prediction ($ \Delta a_\mu=a_\mu^\text{Exp}-a_\mu^\text{SM}=(249\pm49)\times 10^{-11}$)~\cite{Muong-2:2023cdq,Muong-2:2021ojo,Muong-2:2006rrc}, underlying the tendency of new theoretical predictions \cite{Keshavarzi:2018mgv,Colangelo:2017fiz,Blum:2019ugy,Jegerlehner:2009ry} to converge to the observed experimental value $a_{\mu}^{\text{Exp}}$. 

Let us now suppose that the anomalous magnetic moment of the muon is due to some NP; its effects could be manifest in the Lepton Flavor Violating (LFV) decay in the charged lepton sector. 
We consider the $\ell_\alpha\rightarrow\ell_\beta\gamma$ decays, among which the most severe constraint comes from the branching ratio $\mathcal{B}(\mu\to e \gamma)<1.5\times 10^{-13}\, (90\%\text{ C.L.})$ by the MEGII experiment~\cite{MEGII:2025gzr}. Whereas, the current upper-bounds for $\mathcal{B}(\tau\to \mu \gamma)$ and $\mathcal{B}(\tau\to e \gamma)$ are $4.2\times 10^{-8}$and $3.3\times 10^{-8}$, respectively~\cite{BaBar:2009hkt,Belle:2021ysv}. 
Assuming that new degrees of freedom lie above the electroweak scale and can be integrated out, we analyze the electric and magnetic dipole moments of leptons within the framework of the SM Effective Field Theory. In particular, we focus on LFV transitions by exploring two complementary scenarios. In the former case, we impose the current best-fit value on $(g-2)_\mu$, and show that %in our models it is not possible to simultaneously satisfy this constraint 
in our models it is not compatible with the stringent experimental bounds on the decays $\ell_\alpha\rightarrow \ell_\beta \gamma$. In the second scenario, we relax the assumption on $(g-2)_\mu$ and demonstrate that effective operators can account for LFV processes without violating existing experimental limits.
\subsection{The leptonic dipole operator}
If the SM is understood as an EFT valid at the electroweak scale, new higher-dimensional operators $\mathcal{O}^{(n)}_i$ should be considered in the Effective Lagrangian, suppressed by a high-energy scale $\Lambda $ (which is in general a different scale with respect to the cut-off scale introduced in Section~\ref{sec:model}):
\begin{equation}
    \mathcal{L}_\text{SMEFT}=\mathcal{L}_\text{SM}+\left(\sum_{n=1}^d\sum_i \frac{\mathcal{C}^{(n)}_i}{\Lambda^{n-4}}\mathcal{O}^{(n)}_i+\text{H.c.}\right)\,, \quad\text{for} \,\,d>4\,,
\end{equation}
where $n=5,\dots,d$ runs over the dimension of the effective operator labeled by $i=1,\dots$ and $\mathcal{C}^{(n)}_i$ are dimensionless Wilson Coefficients (WCs). Notice that we dropped the flavor structures in the WCs as well as in the effective operators; we will explicitly show them when needed in the following.
Below the electroweak symmetry breaking, the LFV processes are dictated by the dimension-6 leptonic dipole operators given as (in the left-right basis):
\begin{equation}
    \mathcal{O}^{(6)}_{LR}=\frac{v}{\sqrt{2}} \bar{E}_L \sigma^{\mu\nu} E_R F_{\mu\nu}\,,
    \label{eq:dip}
\end{equation}
where $E_L$ and $E_R$ denote three flavors of left-handed and right-handed leptons, respectively. Also, $F_{\mu\nu}$ is the electromagnetic field strength tensor and $v\simeq 246$ GeV denotes the VEV of the Higgs doublet $H$. The operator in eq.~(\ref{eq:dip}) encodes the source of lepton flavor violation that we want to study as arising from our model. Thus, the relevant dipole effective Lagrangian is:
\begin{equation}
    \mathcal{L}_\text{dipole}=\frac{1}{\Lambda^2}\left( \mathcal{C}^\prime_{LR}\mathcal{O}^{(6)}_{LR}+\mathcal{C}^\prime_{RL}\mathcal{O}^{(6)}_{RL}\right)\,, \quad\quad {\mathcal{C}^\prime}^\dagger_{RL}=\mathcal{C}^\prime_{LR}=\begin{pmatrix}
        \mathcal{C}_{ee}^\prime&\mathcal{C}^\prime_{e\mu}&\mathcal{C}^\prime_{e\tau}\\
        \mathcal{C}^\prime_{\mu e}&\mathcal{C}^\prime_{\mu\mu}&\mathcal{C}^\prime_{\mu\tau}\\
        \mathcal{C}^\prime_{\tau e}&\mathcal{C}^\prime_{\tau \mu}&\mathcal{C}^\prime_{\tau\tau}
    \end{pmatrix}\,,
\end{equation}
where the prime of the WCs means that we are working in the mass-eigenstate basis of the charged leptons. The anomalous magnetic moment of the charged lepton $\ell$ can be written at the tree level in terms of the WCs of the dipole operator as follows:
\begin{equation}
    \Delta a_\ell=\frac{4m_\ell}{e}\frac{v}{\sqrt{2}}\frac{1}{\Lambda^2}\Re\left[ \mathcal{C}^\prime_{\ell\ell}\right]\,.
\end{equation}
Then, using the input value $\Delta a_\mu=38\times 10^{-11}$ we obtain:
\begin{equation}
    \frac{1}{\Lambda^2}\Re\left[ \mathcal{C}^\prime_{\mu\mu}\right]=1.5 \times 10^{-6} \text{ TeV}^{-2}\,.
\end{equation}
On the other hand, the tree-level expression of the radiative LFV Branching Ratio (BR) reads:
\begin{equation}
    \mathcal{B}\left( \ell_\alpha\rightarrow\ell_\beta \gamma\right)=\frac{m^3_{\ell_\alpha}\,v^2}{8\pi\Gamma_{\ell_\alpha}} \frac{1}{\Lambda^4}\bigg( |\mathcal{C}^\prime_{\alpha\beta}|^2+|\mathcal{C}^\prime_{\beta\alpha}|^2\bigg)\,,
    \label{eq:LFVBR}
\end{equation}
where $\Gamma_{\ell_\alpha}$ is the total decay width of the lepton $\ell_\alpha$.
Thus, using the current upper bounds from the latest experiments on the LFV processes listed in Table~\ref{refValues}, it is possible to obtain the upper bounds of the Wilson Coefficients. The most stringent bound comes from the $\mu\rightarrow e\gamma$ and it gives:
\begin{equation}
     \frac{1}{\Lambda^2}\Re\left[ \mathcal{C}^\prime_{e\mu(\mu e)}\right]=1.3 \times 10^{-10} \text{ TeV}^{-2}\,,
\end{equation}
while, from branching ratios of the other LFV decays $\tau\rightarrow \mu\gamma$ and $\tau\rightarrow e\gamma$, we obtain:
\begin{equation}
     \frac{1}{\Lambda^2}\Re\left[ \mathcal{C}^\prime_{\mu\tau(\tau\mu)}\right]=2.7 \times 10^{-6} \text{ TeV}^{-2}\,,\quad \frac{1}{\Lambda^2}\Re\left[ \mathcal{C}^\prime_{e\tau(\tau e)}\right]=2.4 \times 10^{-6} \text{ TeV}^{-2}\,,
\end{equation}
respectively.
%=====================TABLE
\renewcommand{\arraystretch}{1.4}
\begin{table}[t]
\centering
\begin{tabularx}{\textwidth}{Clc}
\hline

\bf{Observables} & \bf{Exp.-SM/Bound} & \bf{Wilson Coef. in $1/\Lambda^2$ $[\textbf{TeV}^{-2}$]}\\ 
\hline
$\Delta a_\mu$&$38\times 10^{-11}$~\cite{Muong-2:2025xyk,Muong-2:2023cdq,Muong-2:2021ojo,Muong-2:2006rrc,Aliberti:2025beg}&$\Re \left[C^\prime_{\mu\mu}\right]=1.5\times 10^{-6}$\\
$\mathcal{B}(\mu\rightarrow e\gamma)$&$<1.5\times 10^{-13}$~\cite{MEGII:2025gzr}&$|C^\prime_{e\mu\,(\mu e)}|<1.3\times 10^{-10}$\\
$\mathcal{B}(\tau\rightarrow \mu\gamma)$&$<4.2\times 10^{-8}$~\cite{BaBar:2009hkt,Belle:2021ysv}&$|C^\prime_{\mu\tau\,(\tau\mu)}|<2.7\times 10^{-6}$\\
$\mathcal{B}(\tau\rightarrow e\gamma)$&$<3.3\times 10^{-8}$~\cite{BaBar:2009hkt,Belle:2021ysv}&$|C^\prime_{e\tau\,(\tau e)}|<2.4\times 10^{-6}$\\
\hline
\hline
\end{tabularx}
\caption{Relevant observables and their current upper bounds from the latest experiments. In the last column the corresponding values of the Wilson coefficients presented in $1/\Lambda^2 \left(\text{TeV}^{-2}\right)$ unit.}
\label{refValues}
\end{table}
\renewcommand{\arraystretch}{1.2}
%=====================TABLE
\noindent

In our analysis, we consider the WC to be evaluated at the weak scale, neglecting the small effect of running below that scale~\cite{Buttazzo:2020ibd}.
In the third column of Table~\ref{refValues}, we report the values of the Wilson coefficients (in units of $1/\Lambda^2$ $[\text{TeV}^{-2}]$) corresponding to the upper bounds of the considered LFV processes.

\subsection{Numerical analyses}
Let us discuss the dipole operator in our model. Since in the SMEFT approach the low-energy predictions will not depend on the transformation properties of the flavons under the $U(2)_F$, our analyses will only be about the difference between the A-Scenario and B-Scenario, i.e. we will distinguish the models only by the value of the symmetry breaking parameters $\epsilon_{\phi}$ and $\epsilon_{\chi}$.

The flavor structure of the dipole operator is the same as the Yukawa matrix in eq.~(\ref{eq:yuke}), which is diagonalized by the unitary transformation $U_L^\dagger Y U_R$. 
The unitary matrices can be parametrized as 
\begin{equation}
    U_{L(R)}=U_{L(R)}^{23}\,U_{L(R)}^{13}\,U_{L(R)}^{12}\,,
\end{equation}
and the rotations are given by:
\begin{equation}
    U^{12}=\begin{pmatrix}
        c_{12}&s_{12}&0\\
        -s_{12}&c_{12}&0\\
        0&0&1
    \end{pmatrix}\,,\quad U^{13}=\begin{pmatrix}
        c_{13}&0&s_{13}\\
        0&1&0\\
        -s_{13}&0&c_{13}
    \end{pmatrix}\,,\quad U^{23}=\begin{pmatrix}
        1&0&0\\
        0&c_{23}&s_{23}\\
        0&-s_{23}&c_{23}
    \end{pmatrix}\,,
\end{equation}
where $c_{ij}=\cos\theta_{ij}$ and $s_{ij}=\sin\theta_{ij}$. Also, notice that in general the mixing matrices $U_{L,R}$ would contain three phases (one for each rotation). However, since in our model the charged lepton Yukawa matrix is chosen to be real, such phases are irrelevant (inserting a CP-violation source in the charged lepton sector and then considering non-zero phases in the $U_{L,R}$ matrices would not change our results.\footnote{We are not discussing the Electric Dipole Moment (EDM)~\cite{Roussy:2022cmp,ACME:2018yjb,Kara:2012ay,ACME:2016sci,Muong-2:2008ebm,Belle:2002nla,Bernreuther:2021elu}, therefore our results do not depend on the imaginary part of the Wilson Coefficients. A detailed analysis on the EDM is beyond the scope of this paper and we leave it for a possible future project.})

In the mass-eigenstate basis of the charged leptons, the Wilson Coefficients associated with the dipole operator will be $\mathcal{C}^\prime_{\alpha\beta}=(U_L^\dagger \,\mathcal{C} \,U_R)_{\alpha\beta}$. The predictions on the LFV processes will depend on the flavor structure of the dipole operator which directly originates from the $U(2)_F$ flavor symmetry, from the unitary matrices $U_{L,R}$ which are determined according to the best-fit analyses presented in Sec.~\ref{sec:fit} and from the numerical values of the Wilson Coefficient, that must to be understood as free parameters of the theory. In our numerical analyses the latter are extracted uniformly in the $[\lambda,\lambda^{-1}]$ range. Also, for the sake of simplicity, we assume them to be real, since our results will not depend on the imaginary part of the WCs.
\subsubsection{Predictions of the $(g-2)_{e,\tau}$}

\begin{figure}[t]
    \centering
\includegraphics[width=.9\linewidth]{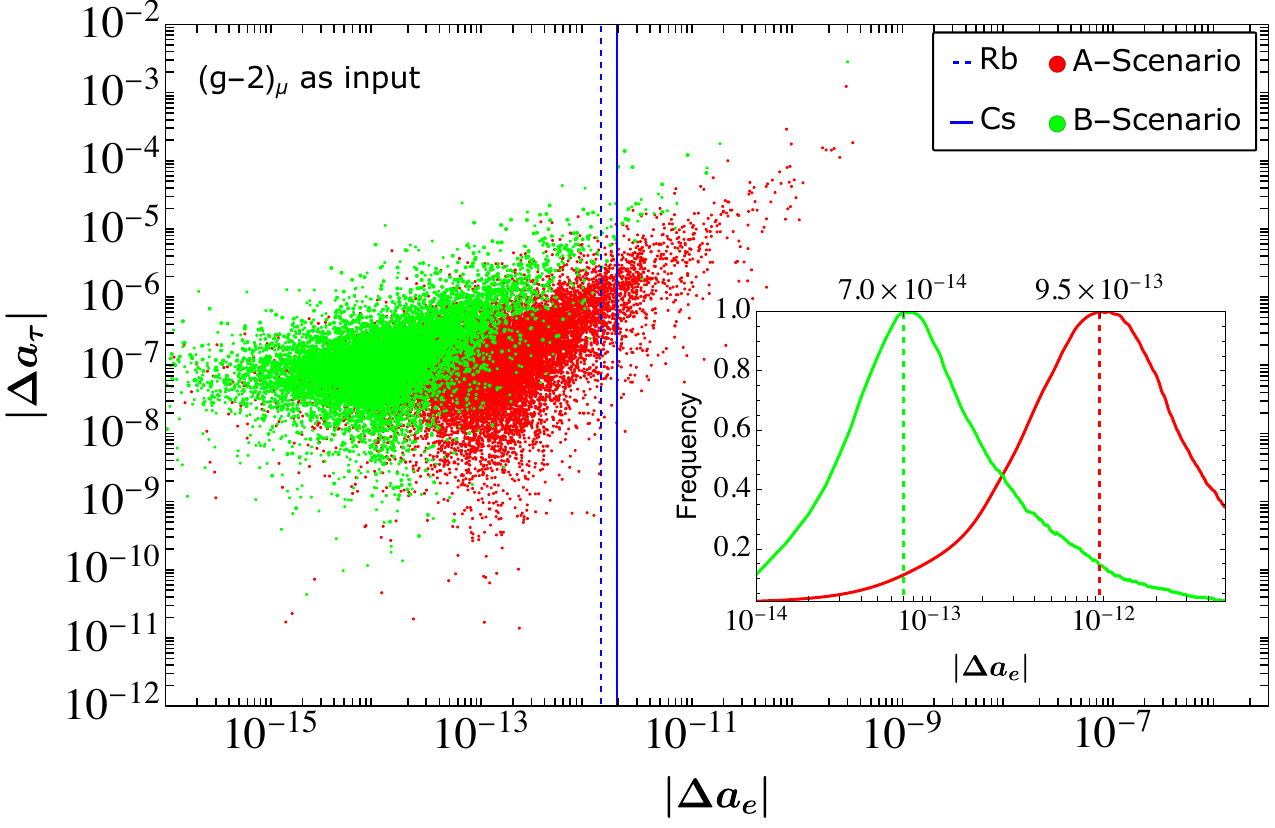}
    \caption{Correlations among $|\Delta a_e|$ and $|\Delta a_\tau|$ arising in the A-Scenario (B-Scenario) and represented with red (green) points,
    obtained considering the reference value for the anomalous magnetic moment as listed in Tab.~\ref{refValues}. The numerical analysis is done assigning random values to the Wilson Coefficients in the $[\lambda,\lambda^{-1}]$ interval. In solid (dashed) blue line the 3$\sigma$ bound for the expected deviation from the SM prediction of the electron magnetic moment according to the latest observations based on Cesium (Rubidium) atomic recoil experiments. 
    In the inset, with the same color code, we show the frequency distributions of the value of $\Delta a_e$ in A-Scenario and B-Scenario. See the main text for further details.}
    \label{fig:putting}
\end{figure}\noindent
Let us assume that the dipole operator is responsible for the observed $(g-2)_\mu$. Thus, we impose the input value listed in Tab.~\ref{refValues} for the $\mathcal{C}_{\mu\mu}^\prime$ and we discuss the electron and the $\tau$ anomalous magnetic moments in our models.

The latest measurements of $(g-2)_{e,\tau}$~\cite{Hanneke:2008tm} give
\begin{equation}
    a_e^{\text{exp}}=1159652180.73(28)\times 10^{-12}\,,
\end{equation}
and this must be compared to the SM prediction which strongly depends on the input value of the fine-structure constant $\alpha$. The latest two measurements based on the Rubidium~\cite{Morel:2020dww} and Cesium~\cite{Parker:2018vye} atomic recoil experiments are in a significant tension, despite being both compatible with the SM prediction within $1\sigma$ and $2.5\sigma$, respectively:
\begin{equation}
    \begin{aligned}
        \Delta a_e^{\text{Rb}}=a_e^{\text{exp}}-a_e^{\text{SM,Rb}}=&\,(+4.8\pm 3.0)\times 10^{-13}\,,\\
        \Delta a_e^{\text{Cs}}=a_e^{\text{exp}}-a_e^{\text{SM,Cs}}=&\,(-8.8\pm 3.6)\times 10^{-13}\,.
    \end{aligned}
    \label{RbCs}
\end{equation}
The aforementioned tension arises mostly due to their opposite signs. However, our model does not favor one sign over the other and, for this reason, we will focus only on the absolute value of the electron magnetic moment. 
The ratios of the diagonal components of the Wilson Coefficients in the A-Scenario and B-Scenario are\footnote{For the analytical expansion of the unitary mixing matrices $U_{L,R}$ in terms of the charged leptons masses we refer to the results in Ref.~\cite{Falkowski:2015zwa}.}
(up to $\mathcal{O}\left(\lambda^4 \,\sqrt{m_e/m_\mu}\right)$):
\begin{equation}
\begin{aligned}
    \textbf{A-Scenario}\,,\quad&\left| \frac{\mathcal{C}^\prime_{ee}}{\mathcal{C}^\prime_{\mu\mu}}\right|\sim \frac{m_e}{m_\mu}\left(1+2\sqrt{\frac{m_\mu}{m_e}}\frac{\sqrt{\cos{\theta_{23}^L}}}{\sin{\theta_{23}^R}}\frac{\mathcal{C}_{e\mu}\lambda^2}{\left(\cos{\theta_{23}^L\mathcal{C}_{\mu\tau}-\sin{\theta_{23}^R\mathcal{C}_{\tau\tau}}}\right)}\right)\,,\\
    \textbf{B-Scenario}\,,\quad&\left| \frac{\mathcal{C}^\prime_{ee}}{\mathcal{C}^\prime_{\mu\mu}}\right|\sim\frac{m_e}{m_\mu}\left(1-\frac{\left(1-\cos{\theta_{23}^R}\right)\mathcal{C}_{\mu\mu} \lambda^2}{\sin{\theta_{23}^R}\mathcal{C}_{\tau\tau}} \right)\,,
    \end{aligned}
\end{equation}
where $\theta_{23}^L$ and $\theta_{23}^R$ are the mixing angles contained in the $U_L$ and $U_R$ unitary matrices, respectively.
Both the A- and B-Scenarios agree on the prediction of the ratio $|\mathcal{C}^\prime_{ee}/\mathcal{C}^\prime_{\mu\mu}|\sim m_e/m_\mu$ at the Leading Order (LO), but it is evident that the Next-To-Leading Order (NLO) contribution in the A-Scenario is enhanced by a multiplicative factor $\sqrt{m_\mu/m_e}$. This implies that the A-Scenario predicts a generally bigger $|\Delta a_e|$ compared to the B-Scenario. 
This is evident in Fig.~\ref{fig:putting}, where the $|\Delta a_e|\text{ vs }|\Delta a_\tau|$ is shown. We performed a scan on the WCs extracting their modulus flat in $[\lambda,\lambda^{-1}]$, with the red (green) points referring to A-Scenario (B-Scenario) predictions. 
The solid (dashed) vertical blue lines refer to the $3\sigma$ upper bounds for the $|\Delta a_e|$ from the latest experiments based on Cesium (Rubidium) atomic recoils. 
It is evident that the B-Scenario predicts a smaller value of $|\Delta a_e|$, while the predictions of $|\Delta a_\tau|$ do not manifest any difference between the two classes of models. 
In order to understand what the preferred predicted value of $\Delta a_{e}$ in each scenario is, we show its frequency distribution in the inset plot of Fig.~\ref{fig:putting}. We normalized the frequency distribution to 1  and show it for both A and B Scenarios, in red and green, respectively. The values corresponding to the peak of the distributions are:
\begin{equation}
    \begin{aligned}
        \textbf{\small{\textsc{A-Scenario}}}:\,\quad &\Delta a_e=\frac{4 m_e}{e}\frac{v}{\sqrt{2}}\frac{1}{\Lambda^2}\Re[\mathcal{C}^\prime_{ee}]\simeq 7.0\times 10^{-14}\,,\\
        \textbf{\small{\textsc{B-Scenario}}}:\,\quad &\Delta a_e=\frac{4 m_e}{e}\frac{v}{\sqrt{2}}\frac{1}{\Lambda^2}\Re[\mathcal{C}^\prime_{ee}]\simeq 9.5\times 10^{-15}\,.
    \end{aligned}
\end{equation}
It is worth to mention that the predicted values are inside the current $3\sigma$ upper bounds and, for this reason, it is not possible to test our models of flavor at present. A similar analysis for  $\Delta a_\tau$ gives:
\begin{equation}
    \Delta a_\tau=\frac{4 m_\tau}{e}\frac{v}{\sqrt{2}}\frac{1}{\Lambda^2}\Re[\mathcal{C}^\prime_{\tau\tau}]\simeq 8.9\times 10^{-8}\,,
\end{equation}
for both the scenarios.
Due to the lack of more precise measurements, we cannot use this prediction to test our models.

Notice that, for both the electron and tau anomalous magnetic moments, our results are approximately in agreement with the naive scaling $\Delta a_\ell\approx m_\ell^2$~\cite{Giudice:2012ms}, even though the presence of the additional $U(2)_F$ flavor symmetry introduces slight deviations.
\subsubsection{LFV decays $\ell_\alpha\rightarrow \ell_\beta\gamma$ in light of $(g-2)_\mu$}
\label{LFVinlightofg-2}
\begin{figure}[t]
    \centering
\includegraphics[width=.9\linewidth]{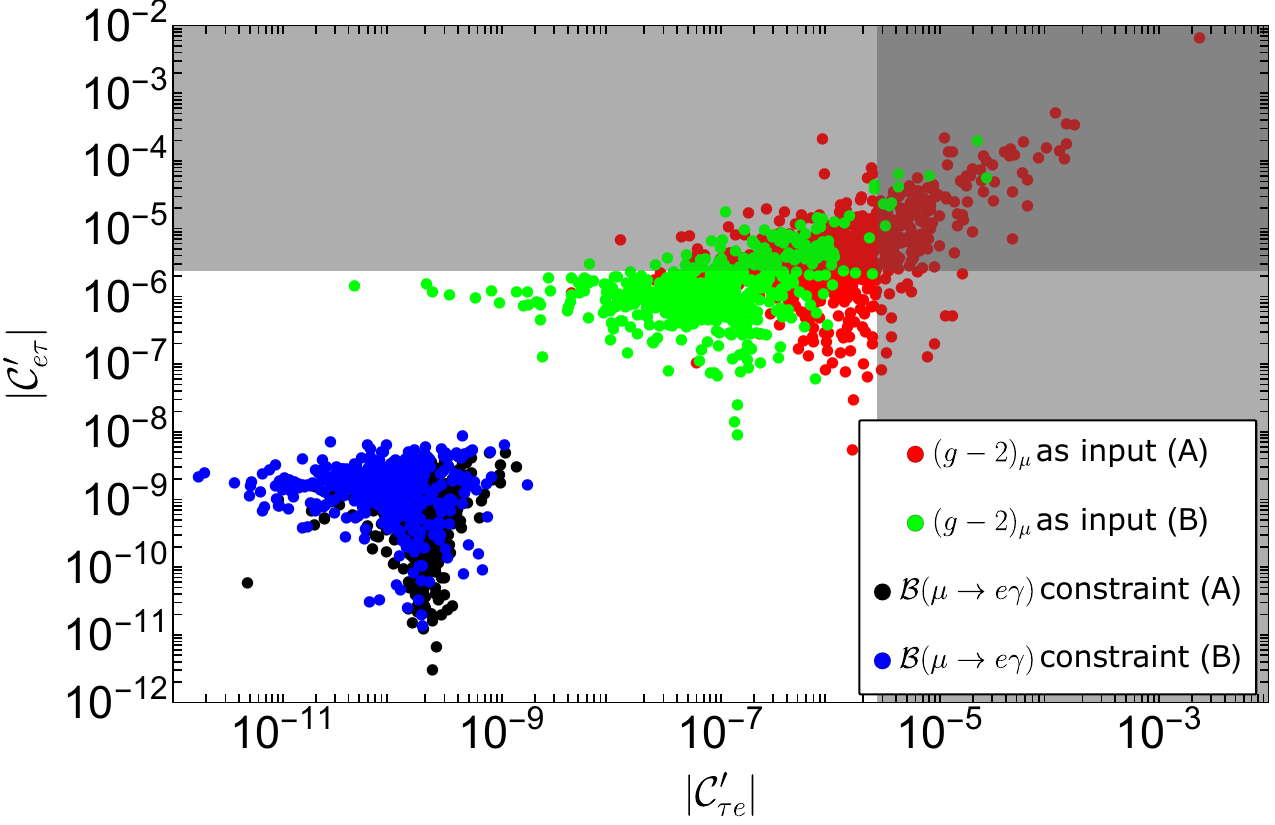}
    \caption{$\big| \mathcal{C}^\prime_{e\tau}\big|$ in $1/\Lambda^2[\text{TeV}^{-2}]$ unit versus $\big| \mathcal{C}^\prime_{\tau e}\big|$. The dark grey region is excluded by the experiments. In red (green) the predictions in the A-Scenario (B-Scenario) in light of the $(g-2)_\mu$. In black (blue) the predictions in the A-Scenario (B-Scenario) relaxing the requirement to reproduce the current $(g-2)_\mu$ value and using the current upper bound $\mathcal{B}(\mu\rightarrow e\gamma)$ as a constraint.}
    \label{fig:Ceτ}
\end{figure}\noindent
The NP responsible for the slight deviation of the central value of the muon's magnetic moment with respect to the SM prediction could manifest in LFV decays. In the following, we will investigate our models predictions of LFV decays, whose most severe bound comes from the MEGII experiment $\mathcal{B}(\mu\rightarrow e\gamma)<1.5\times 10^{-13}$~\cite{MEGII:2025gzr}, in light of the $(g-2)_\mu$,

The relevant WC ratios are:
\begin{enumerate}[label=\bf{\small\textsc{\Alph*-Scenario:}}]
\item $ $\\
\begin{equation}
\begin{aligned}
    \bigg| \frac{\mathcal{C}^\prime_{e\mu}}{\mathcal{C}^\prime_{\mu\mu}}\bigg|\sim&\frac{1}{\sqrt{\cos\theta_{23}^L}} \sqrt{\frac{m_e}{m_\mu}}+\cot{\theta_{23}^R}\frac{\mathcal{C}_{e\mu}\lambda^2}{\cos{\theta_{23}^L}\mathcal{C}_{\mu\tau}-\sin{\theta_{23}^L}\mathcal{{C}_{\tau\tau}}}+\mathcal{O}(\lambda^4)\,,\\
    \bigg| \frac{\mathcal{C}^\prime_{\mu e}}{\mathcal{C}^\prime_{\mu\mu}}\bigg|\sim &\sqrt{\cos\theta_{23}^L}\sqrt{\frac{m_e}{m_\mu}}+\cot{\theta_{23}^R}\frac{\mathcal{C}_{e\mu}\lambda^2}{\cos{\theta_{23}^L}\mathcal{C}_{\mu\tau}-\sin{\theta_{23}^L}\mathcal{{C}_{\tau\tau}}}+\mathcal{O}(\lambda^2\sqrt{m_e/m_\mu})\,;
    \end{aligned}
    \label{eq:cemuA}
\end{equation}
$ $\\
\item $ $\\
\begin{equation}
\begin{aligned}
    \bigg| \frac{\mathcal{C}^\prime_{e\mu}}{\mathcal{C}^\prime_{\mu\mu}}\bigg|\sim&\frac{1}{\sqrt{\cos\theta_{23}^L}} \sqrt{\frac{m_e}{m_\mu}}+\cot{\theta_{23}^R}\frac{\mathcal{C}_{e\mu}\lambda^2}{\cos{\theta_{23}^L}\mathcal{C}_{\mu\tau}-\sin{\theta_{23}^L}\mathcal{{C}_{\tau\tau}}}+\mathcal{O}(\lambda^4)\,,\\
    \bigg| \frac{\mathcal{C}^\prime_{\mu e}}{\mathcal{C}^\prime_{\mu\mu}}\bigg|\sim &\sqrt{\cos\theta_{23}^L}\sqrt{\frac{m_e}{m_\mu}}+\cot{\theta_{23}^R}\frac{\mathcal{C}_{e\mu}\lambda^2}{\cos{\theta_{23}^L}\mathcal{C}_{\mu\tau}-\sin{\theta_{23}^L}\mathcal{{C}_{\tau\tau}}}+\mathcal{O}(\lambda^2\sqrt{m_e/m_\mu})\,.
    \end{aligned}
    \label{eq:cemuB}
\end{equation}
\end{enumerate}
Our models do not manifest a suppression of $\mathcal{C}^\prime_{e\mu}$ compared with $\mathcal{C}^\prime_{\mu e}$. It means that even though the angular distribution with respect to the muon polarization can distinguish between $\mu_R\rightarrow e_L\gamma$ and $\mu_L\rightarrow e_R\gamma$~\cite{Okada:1999zk}, this cannot be used to probe our models at present.
Also, it is evident that assuming all the WCs to be of $\mathcal{O}(1)$, for both the A-Scenario and the B-Scenario the $\mathcal{C}^\prime_{e\mu (\mu e)}$ is suppressed by at most an $\mathcal{O}(10^{-2})$ with respect to $\mathcal{C}^\prime_{\mu\mu}$, which implies that $\mathcal{C}^\prime_{e\mu(\mu e)}\sim\mathcal{O}(10^{-8})$. Therefore, following eq.~(\ref{eq:LFVBR}), the predicted $\mathcal{B}(\mu\rightarrow e\gamma)$ violates the current experimental upper bound listed in Table~\ref{refValues} by $2\div3$ orders of magnitude. This is not surprising because the same NP couples similarly to the electron and muon BSM sectors so that\footnote{Many works attempted to disentangle the muon and electron sector in order to simultaneously address the $(g-2)_\mu$ anomaly and the $\mu\rightarrow e \gamma$ bounds, see for instance ref.~\cite{Crivellin:2019mvj,Dermisek:2021ajd,Crivellin:2020tsz} and reference therein.}
\begin{equation}
    \mathcal{B}(\mu\rightarrow e\gamma)=\dfrac{\alpha m_\mu^2}{16m_e\Gamma_\mu} \big| \Delta a_\mu \Delta a_e\big|\,,
    \label{aeamuBR}
\end{equation}
where $\alpha$ is the fine-structure constant and $\Gamma_\mu$ the total decay width of the muon.
Adopting for $\Delta a_\mu$ the current central value (see Table~\ref{refValues}) as our reference point, we find that this value is not compatible with the current bounds on $\mu\rightarrow e\gamma$, unless a certain degree of fine tuning in the Wilson coefficients is assumed.

\begin{figure}[t]
    \centering
\includegraphics[width=.9\linewidth]{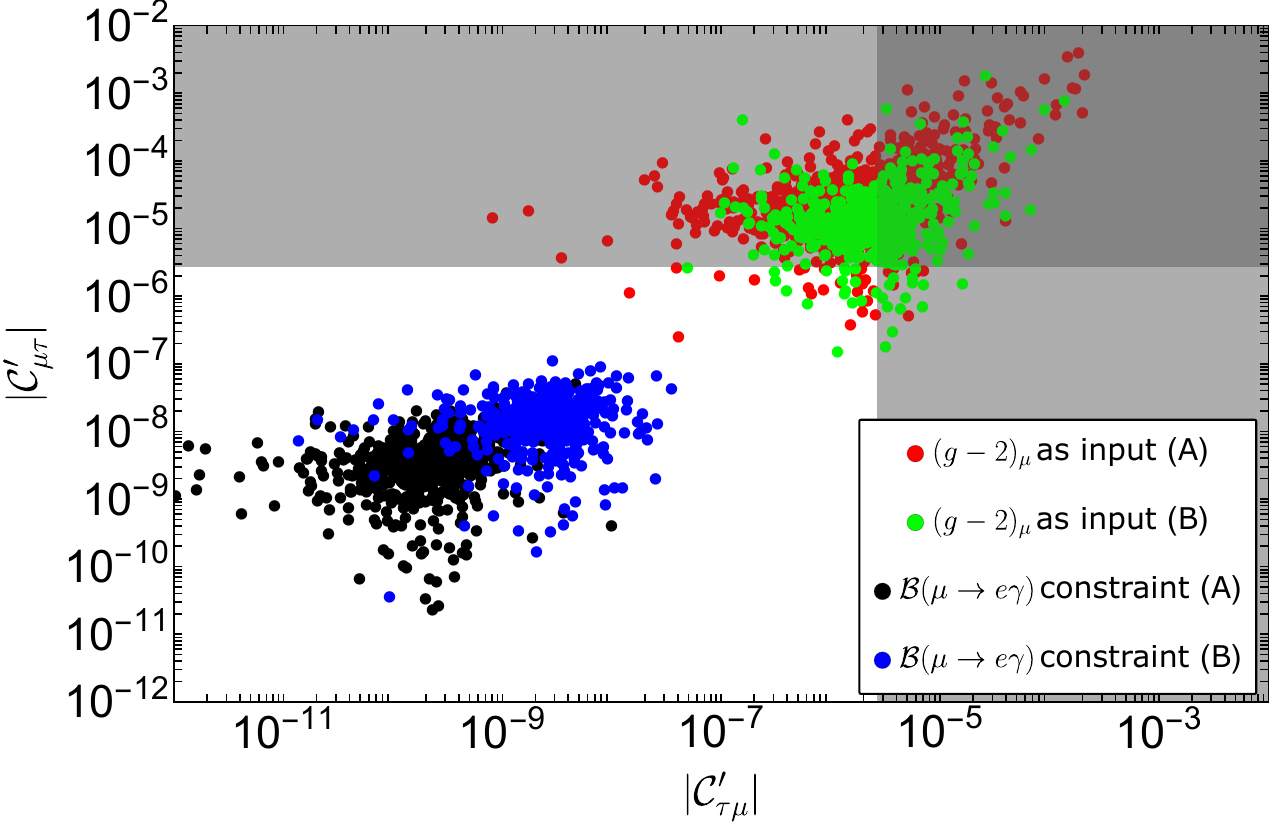}
    \caption{$\big| \mathcal{C}^\prime_{e\tau}\big|$ in $1/\Lambda^2[\text{TeV}^{-2}]$ unit versus $\big| \mathcal{C}^\prime_{\tau e}\big|$. The dark grey region is excluded by the experiments. The color code is the same as of Fig.~\ref{fig:Ceτ}.}
    \label{fig:Cμτ}
\end{figure}

Also, the predicted BRs of the radiative LFV decays $\tau\rightarrow e\gamma$ and $\tau\rightarrow\mu\gamma$ are in mild tension with the latest upper-bounds: $\mathcal{B}(\tau\to e \gamma)<4.2\times 10^{-8}$ and $\mathcal{B}(\tau\to \mu \gamma)<3.3\times 10^{-8}$~\cite{BaBar:2009hkt,Belle:2021ysv}. To keep the text readable, we leave the analytical expressions of the associated Wilson Coefficient ratios in Appendix~\ref{ap:Cmutau} and we only present here the numerical results in Figures~\ref{fig:Ceτ} and~\ref{fig:Cμτ}. The absolute values of the relevant WCs in the A-Scenario and B-Scenario are shown in red and green circles, respectively. 

%\sm{mitigate this, in the sense that at 1$\sigma$ the tension between the $g-2$ and the $\mu\rightarrow e\gamma$ disappears. }It is evident that using the value of the $(g-2)_\mu$ anomaly as input, all models violate the latest upper bounds of the LFV decays involving tauons (listed in Table~\ref{refValues}). 
We can relax the requirement to reproduce the anomalous magnetic moment of the muon and impose the condition that our models satisfy the most severe bound of $\mu\rightarrow e\gamma$. The result will be that the predicted $(g-2)_{e,\mu}$ are strongly suppressed according to the expected relation in eq.~(\ref{aeamuBR}) and the SM predictions will be essentially recovered. In addition, the other LFV decays $\tau\rightarrow e\gamma$ and $\tau\rightarrow\mu\gamma$ rates will be predicted according to the latest upper bounds. This is shown in Figures~(\ref{fig:Ceτ},\ref{fig:Cμτ}) where the black and blue circles represent the relevant absolute values of the WCs imposing the constraint $\mathcal{B}(\mu\rightarrow e\gamma)$ in the A-Scenario and B-Scenario, respectively. \\
It is important to highlight that, within an $U(2)_F$ flavor model, the current experimental limits on the branching ratios of LFV decays of the $\tau$ leptons are only marginally compatible with the observed $\Delta a_\mu$, while the bounds on $\mu\rightarrow e\gamma$ are too tight to allow even a slight anomalous magnetic moment of the muon. This suggests that the current central value of the $(g-2)_\mu$ deviation from the Standard Model, despite being reduced by a factor of 8 by the latest calculations, still cannot be consistently explained unless the electron and muon sectors are effectively decoupled. This can be accomplished, for instance, by invoking UV completions. However, this is beyond the scope of this paper and we leave it for a possible follow-up study.

\section{Conclusions}
\label{conclusions}
The discovery of neutrino oscillations, and hence the fact that neutrinos are massive, has introduced a wide range of unanswered questions into particle physics. For this reason, it has become of fundamental importance to construct a flavor symmetry capable not only of predicting the mass spectrum across generations, but also of accounting for the mixing patterns of both neutrinos and quarks, which exhibit markedly different features. In this work, we wanted to extend to the neutrino sector the model presented in \cite{Linster:2018avp} where an $U(2)_F$ flavor symmetry, isomorphic to $SU(2)_F\times U(1)_F$, has been proposed to predict quark and leptons masses and mixings with the introduction of two flavons: an $SU(2)_F$ doublet $\phi$ and a singlet $\chi$. Such a model failed to recover the observed neutrino mass spectrum via type-I see-saw when some assumptions on the quantum numbers of Majorana neutrinos under the flavor symmetry were made. We show that: a) relaxing the requirement that the $SU(2)_F$ structure of Majorana neutrino should reflect the $SU(2)_F$ structure of lepton doublets and b) allowing for them to assume also negative $U(1)_F$ charges, it is possible to fit all neutrinos and leptons observables successfully. In order to make a complete classification of all possible charge assignments, we distinguished three different classes of Models, depending on the $SU(2)_F$ structure of Majorana neutrinos:
\begin{itemize}
    \item Model \textbf{S}: the three Majorana neutrinos are singlets under $SU(2)_F$
    \item Model \textbf{D}: two Majorana neutrinos ($N_1$ and $N_2$) are in a flavor doublet while $N_3$ is a singlet
    \item Model \textbf{T}: the three Majorana neutrinos transform as a flavor triplet.
\end{itemize}
Successively, using the see-saw type-I mechanism, we searched for all the possible light neutrino mass matrices scanning over the $U(1)_F$ charges of Majorana neutrinos. By considering mass matrix patterns as distinct when the difference between corresponding entries was at least of order $\lambda^4$ (with $\lambda\sim0.2$) smaller than the leading term of that entry, we identified a total of 104 different patterns. We further classified them in A and B Scenarios according on whether the flavons VEVs were considered of the same order of magnitude or hierarchical (that is, $\varepsilon_\chi/\varepsilon_\phi\sim\lambda$), respectively. Performing a fit on the neutrino observables (mixing angles and $\Delta m_{ij}^2$) as well as on the ratios of charged leptons masses, we found 13 patterns which  reproduce all the observables in a satisfactory way with very low $\chi^2$ values and with a relatively small fine-tuning, estimated through the $d_{FT}$ parameter. 
For all unknown couplings, we left the modulus to vary in the range [$\lambda$, $\lambda^{-1}$] and the phases in the whole $[0,2\pi)$ range.
Out of the 13 patterns, 6 were part of Model \textbf{S} (with two of them belonging to the B-Scenario) while 7 were part of the Model \textbf{D} (with 3 being of the B-Scenario category). No Model \textbf{T} patterns succeeded in the fit, due to small number of free parameters which introduced strong, unbreakable correlations among observables.
The main conclusions derived from a detailed analysis of the 13 patterns are:
\begin{itemize}
    \item All the viable patterns could reproduce only the Normal Ordering.
    \item With some particular choices of charges, anarchical neutrino mixing matrices were found. These particular patterns are known to be able to fit neutrino observable.
    \item The tentative choice for Majorana charges made in \cite{Linster:2018avp} was discarded since, from analytical expressions, it could not reproduce the neutrino mass spectrum. However, their charge choice appeared in one of our 13 viable patterns. We demonstrated that, in this particular case, it is still possible to correctly reproduce the value of $\alpha=\Delta m_{21}^2/\Delta m^2_{31}$ by lowering only one parameter of the Majorana mass matrix to a value similar to $\lambda$. This feature has been observed also in other two viable patterns. The main consequence of this is the lowering of the predicted lightest neutrino mass down to $10^{-7}$ eV.
    \item All the viable patters predict an effective Majorana mass parameter $m_{\beta\beta}$ between $10^{-3}$ and $10^{-2}$ eV, which might be close to the reaches of future $0\nu\beta\beta$ experiments. On the other hand, the predicted effective neutrino mass $m_\beta$ is too low (less than 20 meV) to be reachable by current and next-generation experiments involving tritium decays or Holmium electron capture. 
\end{itemize}
We further observed that, given the presence of unbounded and uncorrelated phases, our model could not predict any specific value of the PMNS  CP-violating phase $\delta$. \\
In the final part of our study, we performed an analysis on the leptonic dipole operators and on the LFV decays within the 13 selected patters. In particular, we assumed all the new physics mediating the processes to be integrated out and, in an EFT approach, we studied the predictions of our $U(2)_F$
 models. We found out that, if we assume the (currently small) $(g-2)_\mu$ anomaly as input for our analysis then the predictions on the LFV process $\mu\to e\gamma$ are not compatible with current measurements. However, if we set the current bounds on $\mu\to e\gamma$ as input, we lower the prediction on $(g-2)_\mu$, setting consistent bounds on other processes like $\tau\to e\gamma$, $\tau\to\mu\gamma$ and on the electron and $\tau$ dipole moment. 
 Therefore, the current tendency of refined theoretical calculation~\cite{Keshavarzi:2018mgv,Colangelo:2017fiz,Blum:2019ugy,Jegerlehner:2009ry,Aliberti:2025beg} and updated experimental analyses \cite{Muong-2:2015xgu,Muong-2:2023cdq,Muong-2:2025xyk} to reduce the tension in the $(g-2)_\mu$ anomaly is completely in line with the prediction of our simple $U(2)_F$ flavor model, which in fact could offer a fully consistent framework not only for explaining neutrino masses and mixing, but also for accommodating new physics capable of mediating LFV processes.

\section*{Acknoledgements}
SM thanks C. Hagedorn for useful discussions and comments. SM acknowledges financial support from the project \emph{"consolidaci\`on investigadora"} CNS2022-13600.
\newpage
\appendix
\section{$(g-2)_\mu$ versus the LFV processes $\tau\rightarrow e\gamma$ and $\tau\rightarrow\mu\gamma$} \label{ap:Cmutau}
In Sec.~\ref{LFVinlightofg-2} we used the analytical expansions in eqs.~(\ref{eq:cemuA},\ref{eq:cemuB}) to affirm that, by using the $(g-2)_\mu$ value as an input for our numerical analysis, the upper-bound for the $\mu\rightarrow e\gamma$ is violated. In this Appendix we want to follow a similar reasoning for the LFV radiative decays of the tauons. Therefore, we present the correlations between the muon anomalous magnetic moment and the LFV radiative processes $\tau\rightarrow e\gamma$ and $\tau\rightarrow\mu\gamma$. 

The ratios of the relevant WC are:
\begin{enumerate}[label=\bf{\small\textsc{\Alph*-Scenario:}}]
\item $ $\\
\begin{equation}
\begin{aligned}
    \bigg| \frac{\mathcal{C}^\prime_{e\tau}}{\mathcal{C}^\prime_{\mu\mu}}\bigg|\sim&\frac{\cot{\theta_{23}^R}}{\sqrt{\cos{\theta_{23}^L}}}\sqrt{\frac{m_e}{m_\mu}}+\frac{\mathcal{C}_{e\mu}\lambda^2}{\kappa}+\mathcal{O}(\lambda^4)\,,\\[5pt]
    \bigg| \frac{\mathcal{C}^\prime_{\tau e}}{\mathcal{C}^\prime_{\mu\mu}}\bigg|\sim &\sqrt{\cos{\theta_{23}^L}}\sqrt{\frac{m_e}{m_\mu}}\left(\frac{\sin{\theta_{23}^L}\mathcal{C}_{\mu\tau}+\cos{\theta_{23}^L}\mathcal{C}_{\tau\tau}}{\kappa}\right)-\frac{\sin{\theta_{23}^L\mathcal{C}_{e\mu}\lambda^2}}{\sin{\theta_{23}^R}\kappa}+\mathcal{O}\left(\lambda^2\sqrt{m_e/m_\mu}\right)\,,\\[5pt]
    \bigg| \frac{\mathcal{C}^\prime_{\mu \tau}}{\mathcal{C}^\prime_{\mu\mu}}\bigg|\sim &\cot{\theta_{23}^R}+\left(\frac{\cos{\theta_{23}^L}\mathcal{C}_{\mu\mu}-\sin{\theta_{23}^L}\mathcal{C}_{\tau\mu}}{{\sin^2{\theta_{23}^R}}\kappa}\right)\lambda^2+\mathcal{O}\left(m_e/m_\mu\right)\,,\\[5pt]
    \bigg| \frac{\mathcal{C}^\prime_{\tau \mu}}{\mathcal{C}^\prime_{\mu\mu}}\bigg|\sim &\frac{\sin{\theta_{23}^L}\mathcal{C}_{\mu\tau}+\cos{\theta_{23}^L}\mathcal{C}_{\tau\tau}}{\kappa}+\cot{\theta_{23}^R}\left(\frac{\mathcal{C}_{\mu\tau}\mathcal{C}_{\tau\tau}-\mathcal{C}_{\mu\mu}\mathcal{C}_{\tau\tau}}{\kappa^2}\right)\lambda^2+\mathcal{O}\left(m_e/m_\mu\right)\,,
    \end{aligned}
    \label{eq:Ataus}
\end{equation}
\item $ $\\
\begin{equation}
\begin{aligned}
    \bigg| \frac{\mathcal{C}^\prime_{e\tau}}{\mathcal{C}^\prime_{\mu\mu}}\bigg|\sim&\frac{\cot{\theta_{23}^R}}{\sqrt{\cos{\theta_{23}^L}}}\sqrt{\frac{m_e}{m_\mu}}+\mathcal{O}\left(\lambda^2\sqrt{m_e/m_\mu}\right)\,,\\[5pt]
    \bigg| \frac{\mathcal{C}^\prime_{\tau e}}{\mathcal{C}^\prime_{\mu\mu}}\bigg|\sim &\sqrt{\cos{\theta_{23}^L}}\tan{\theta_{23}^L}\sqrt{\frac{m_e}{m_\mu}}+\mathcal{O}\left(\lambda \sqrt{m_e/m_\mu}\right)\,,\\[5pt]
    \bigg| \frac{\mathcal{C}^\prime_{\mu \tau}}{\mathcal{C}^\prime_{\mu\mu}}\bigg|\sim &\cot{\theta_{23}^R}+\frac{\mathcal{C}_{\mu\mu}\,\lambda^2}{{\sin^2{\theta_{23}^R}}\mathcal{C}_{\mu\tau}}+\mathcal{O}(\lambda^3)\,,\\[5pt]
    \bigg| \frac{\mathcal{C}^\prime_{\tau \mu}}{\mathcal{C}^\prime_{\mu\mu}}\bigg|\sim &\tan{\theta_{23}^L}+\frac{\mathcal{C}_{\tau\tau}\,\lambda\,\left( \cos{\theta_{23}^L\mathcal{C}_{\mu\tau}+\sin{\theta_{23}^L}\mathcal{C}_{\tau\tau}\,\lambda}\right)}{\mathcal{C}_{\mu\tau}^2}+\mathcal{O}(\lambda^3)\,.
    \end{aligned}
    \label{eq:Btaus}
\end{equation}
\end{enumerate}
where $\kappa=(\sin{\theta_{23}^L}\mathcal{C}_{\tau\tau}-\cos{\theta_{23}^L}\mathcal{C}_{\mu\tau})$. The analytical expansions in eqs.~(\ref{eq:Btaus},\ref{eq:Ataus}) prove that, if we consider the $(g-2)_\mu$ as an input value, our models violate the latest upper-bounds (see Table~\ref{refValues}) for the radiative LFV decays $\tau\rightarrow e\gamma$ and $\tau\rightarrow\mu\gamma$, as it is confirmed by the numerical analysis in Figures~\ref{fig:Ceτ} and~\ref{fig:Cμτ}.
\newpage
\section{Fit results} \label{ap:tab}
The following two tables contain the best-fit parameters for the 13 viable patterns (6 for Model \textbf{S} and 7 for Model \textbf{D}). The complete procedure of the numerical fit is accurately reported in Section \ref{sec:fit}.

\renewcommand{\arraystretch}{.95}
\begin{table}[h]
\centering
%\resizebox{\textwidth}{!}
\begin{tabular}{c|ccccccc}
\Xhline{1.5pt} \rule[-0.27cm]{0pt}{0.8cm}
\textbf{Pattern} & \textbf{S}1 A & \textbf{S}2 A & \textbf{S}3 A & \textbf{S}4 A & \textbf{S}1 B & \textbf{S}2 B \\
\Xhline{1.5pt}

%$\mathbb{R}\text{e}(e_{11})
$e_{11}$ & -0.33 & -3.25 & 2.98 & -2.02 & 3.01 & -4.34 \\
$e_{12}$ & -0.44 & -0.88 & -0.41 & -0.59 & -2.15 & -3.27 \\
$e_{13}$ & 4.09 & 3.81 & -2.20 & 4.13 & 0.33 & -1.04 \\
$e_{22}$ & 4.11 & -0.53 & 2.76 & 2.53 & 1.86 & -2.96 \\
$e_{23}$ & -3.74 & 3.78 & 2.72 & 3.03 & -0.93 & -1.41 \\
$e_{31}$ & 1.23 & 3.63 & 0.60 & 0.91 & 3.71 & 2.29 \\
$e_{32}$ & 2.76 & -3.90 & 3.99 & 3.57 & 1.07 & -1.51 \\
$e_{33}$ & 1.86 & -0.28 & 0.82 & 0.39 & 2.52 & 3.27 \\
\hline
$y_{11}$ & 1.31 & -1.33 & -3.31 & -3.07 & 3.22 & 3.35 \\
$y_{12}$ & 1.73 & -3.55 & -1.60 & -2.02 & -3.68 & -3.95 \\
$y_{13}$ & 1.03 & 2.15 & -4.20 & -4.04 & -0.56 & 0.48 \\
$y_{21}$ & 3.06 & 4.31 & -3.39 & -3.27 & 0.95 & -3.71 \\
$y_{22}$ & -2.41 & 0.48 & 1.40 & -0.38 & 3.85 & 1.53 \\
$y_{23}$ & 2.93 & 1.15 & -1.85 & -2.96 & -0.65 & -2.78 \\
$y_{31}$ & -1.19 & -3.00 & -2.90 & -2.13 & -4.27 & -4.24 \\
$y_{32}$ & 3.66 & -3.01 & -0.52 & 0.41 & 2.77 & 0.36 \\
$y_{33}$ & 0.34 & -1.80 & 1.65 & 4.06 & 2.19 & 1.98 \\
\hline
$k_{11}$ & -1.21 & 1.09 & -4.12 & -1.66 & 4.19 & 0.29 \\
$k_{12}$ & 1.81 & 1.33 & 0.42 & 4.06 & -1.57 & -1.71 \\
$k_{13}$ & -3.89 & 1.94 & -0.24 & 0.78 & 1.66 & -3.52 \\
$k_{22}$ & 3.94 & 3.37 & -1.57 & -1.95 & 1.29 & -2.85 \\
$k_{23}$ & -2.75 & -3.58 & 4.01 & 0.38 & 2.60 & 0.73 \\
$k_{33}$ & 1.09 & 1.55 & -0.54 & -1.35 & -1.82 & -2.89 \\
\hline
$\sin^2{\theta_{12}}$ & 0.302 & 0.302 & 0.302 & 0.308 & 0.307 & 0.305 \\
$\sin^2{\theta_{13}}$ & 0.0222 & 0.0225 & 0.0225 & 0.0219 & 0.0224 & 0.0222 \\
$\sin^2{\theta_{23}}$ & 0.454 & 0.457 & 0.454 & 0.458 & 0.458 & 0.458 \\
$\alpha$ & 0.0294 & 0.0292 & 0.0296 & 0.0294 & 0.0296 & 0.0295 \\
$r_{12}$ & 0.0048 & 0.0048 & 0.0047 & 0.0048 & 0.0048 & 0.0048 \\
$r_{23}$ & 0.0547 & 0.0561 & 0.0567 & 0.0565 & 0.0571 & 0.0563 \\
\hline
$\chi^2$ & 0.24 & 0.45 & 0.64 & 0.70 & 0.34 & 0.23 \\
$d_\text{FT}$ & 7.12 & 4.19 & 12.2 & 6.02 & 5.53 & 5.72 \\
$\chi^2 + \text{P}_{\mathrm{MG}}$ & 27.00 & 26.62 & 26.22 & 25.97 & 24.19 & 29.04 \\
\Xhline{1.5pt}

\end{tabular}
\caption{\label{tab:BFmodS} Best-fit values and fit parameters for Model \textbf{S} viable patterns. For the parameters $e_{ij}$, $y_{ij}$ and $k_{ij}$ we reported only the real part of their best-fit values.}
\end{table}

\begin{table}[h]
\centering
%\resizebox{\textwidth}{!}
\begin{tabular}{c|ccccccc}
\Xhline{1.5pt} \rule[-0.27cm]{0pt}{0.8cm}
\textbf{Pattern} & \textbf{D}1 A & \textbf{D}2 A & \textbf{D}3 A & \textbf{D}4 A & \textbf{D}1 B & \textbf{D}2 B & \textbf{D}5 B \\
\Xhline{1.5pt}

$e_{11}$ & 1.00 & -1.47 & 0.92 & -3.18 & 1.69 & 0.94 & -0.74 \\
$e_{12}$ & -0.32 & -0.53 & 0.34 & -0.31 & -2.69 & 3.38 & -1.28 \\
$e_{13}$ & 0.62 & -1.76 & -2.15 & 3.60 & 2.15 & 0.89 & 1.95 \\
$e_{22}$ & -3.30 & -3.01 & 3.18 & 0.92 & -3.81 & -2.33 & -1.83 \\
$e_{23}$ & -2.67 & 3.10 & -2.24 & -2.40 & 1.19 & -1.40 & 0.50 \\
$e_{31}$ & -2.06 & -1.63 & 3.39 & -3.94 & -1.65 & 4.01 & 3.82 \\
$e_{32}$ & 2.11 & 4.13 & -3.88 & 2.65 & -2.64 & 3.30 & -3.68 \\
$e_{33}$ & -2.23 & -0.65 & -3.97 & 1.06 & 3.20 & -2.77 & 2.61 \\
\hline
$y_{11}$ & 2.36 & -1.76 & 3.30 & 3.74 & 1.64 & 3.02 & 2.97 \\
$y_{12}$ & -1.02 & 0.76 & 2.86 & -2.87 & -2.72 & 3.29 & -1.13 \\
$y_{13}$ & 2.15 & 3.37 & 3.54 & -3.98 & 0.36 & 0.42 & -2.24 \\
$y_{22}$ & -3.05 & 3.89 & -1.14 & 2.47 & -1.61 & -0.75 & 3.20 \\
$y_{23}$ & -1.11 & -1.78 & -2.58 & -3.15 & -2.56 & 2.08 & 3.28 \\
$y_{31}$ & -3.49 & 0.72 & 2.82 & -0.38 & -1.41 & -3.40 & 3.32 \\
$y_{32}$ & 3.44 & -1.59 & 3.79 & 2.50 & 0.86 & 4.32 & 0.88 \\
$y_{33}$ & 2.92 & 2.39 & -0.42 & -1.15 & -3.64 & 1.02 & 1.26 \\
\hline
$k_{11}$ & 1.61 & 3.55 & -1.40 & -1.61 & 3.21 & 4.05 & 2.10 \\
$k_{12}$ & -3.40 & 3.37 & -3.92 & -3.44 & 1.44 & 1.10 & -3.50 \\
$k_{13}$ & -2.55 & -3.41 & 2.31 & 3.76 & 0.39 & -1.80 & 1.60 \\
$k_{22}$ & 2.20 & 4.00 & 3.23 & -1.59 & 1.86 & 3.67 & -0.34 \\
$k_{23}$ & -0.95 & 3.05 & 0.28 & -2.29 & -0.62 & 0.77 & 1.12 \\
$k_{33}$ & 3.65 & 0.56 & 1.82 & -3.49 & 2.16 & -2.11 & -0.25 \\
\hline
$\sin^2{\theta_{12}}$ & 0.297 & 0.305 & 0.305 & 0.297 & 0.300 & 0.305 & 0.303 \\
$\sin^2{\theta_{13}}$ & 0.0221 & 0.0222 & 0.0221 & 0.0219 & 0.0223 & 0.0222 & 0.0222 \\
$\sin^2{\theta_{23}}$ & 0.452 & 0.447 & 0.451 & 0.455 & 0.444 & 0.457 & 0.450 \\
$\alpha$ & 0.0298 & 0.0295 & 0.0296 & 0.0304 & 0.0296 & 0.0298 & 0.0295 \\
$r_{12}$ & 0.0048 & 0.0048 & 0.0048 & 0.0048 & 0.0048 & 0.0048 & 0.0048 \\
$r_{23}$ & 0.0571 & 0.0577 & 0.0546 & 0.0569 & 0.0553 & 0.0570 & 0.0572 \\
\hline
$\chi^2$ & 0.39 & 0.24 & 0.35 & 1.96 & 0.38 & 0.30 & 0.06 \\
$d_\text{FT}$ & 17.1 & 7.03 & 37.6 & 4.26 & 8.41 & 3.31 & 45.0 \\
$\chi^2 + \text{P}_{\mathrm{MG}}$ & 24.44 & 27.71 & 28.10 & 28.43 & 19.97 & 22.33 & 22.07 \\
\Xhline{1.5pt}
\end{tabular}
\caption{\label{tab:BFmodD} Best-fit values and fit parameters for Model \textbf{D} viable patterns. For the parameters $e_{ij}$, $y_{ij}$ and $k_{ij}$ we reported only the real part of their best-fit values.}
\end{table}
\renewcommand{\arraystretch}{1}
\newpage
\bibliographystyle{JHEP}
\bibliography{bibly}
\end{document}